\documentclass{emulateapj}

\pdfoutput=1
\usepackage{epstopdf}
\usepackage{amsmath}
\usepackage{comment}
\usepackage{appendix}
\usepackage{color}

\slugcomment{\today}

\begin{document}
\title{Systematic Uncertainties Associated with the Cosmological Analysis of the First Pan-STARRS1 Type Ia Supernova Sample}

\newcommand{\calnum}{12}
\newcommand{\SNIAPSall}{146}
\newcommand{\SNIAlowztot}{222}
\newcommand{\SNIAtot}{335}
\newcommand{\SNIAPSused}{113}
\newcommand{\SNIAlowzused}{197}
\newcommand{\SNIAused}{310}
\newcommand{\SNIAlowzall}{497}
\newcommand{\wSNstat} {\ensuremath{-1.010^{+0.360}_{-0.206}}}
\newcommand{\wCstat} {\ensuremath{-1.131^{+0.049}_{-0.049}}}
\newcommand{\fomCstat} {\ensuremath{0.284^{+0.010}_{-0.010}}}
\newcommand{\wCsys} {\ensuremath{-1.166^{+0.072}_{-0.069}}}
\newcommand{\fomCsys} {\ensuremath{0.280^{+0.013}_{-0.012}}}
\newcommand{\wCwsys}{\ensuremath{-1.124^{+0.083}_{-0.065}}}
\newcommand{\wSNsys} {\ensuremath{-1.120^{+0.450}_{-0.357}}}
\newcommand{\wSNstasys} {\ensuremath{-1.120^{+0.360}_{-0.206}\textrm{(Stat)} ^{+0.269}_{-0.291}\textrm{(Sys)}}}
\newcommand{\fomSNsys} {\ensuremath{0.256^{+0.201}_{-0.174}}}
\newcommand{\fomSNstat} {\ensuremath{0.223^{+0.209}_{-0.221}}}
\newcommand{\omSNstasys} {\ensuremath{0.217^{+0.263}_{-0.217}\textrm{(Stat)} ^{+-NaN}_{-0.123}\textrm{(Sys)}}}
\newcommand{\snOMstat}{\ensuremath{0.242^{+0.047}_{-0.238}}}
\newcommand{\snOMsys}{\ensuremath{0.242^{+0.047}_{-0.238}}}
\newcommand{\snOLstat}{\ensuremath{0.743^{+0.115}_{-0.252}}}
\newcommand{\snOLsys}{\ensuremath{0.743^{+0.115}_{-0.252}}}
\newcommand{\snOMstatf}{\ensuremath{0.244^{+0.039}_{-0.041}}}
\newcommand{\snOMsysf}{\ensuremath{0.228^{+0.054}_{-0.061}}}
\newcommand{\snOLstatf}{\ensuremath{0.760^{+0.043}_{-0.037}}}
\newcommand{\snOLsysf}{\ensuremath{0.776^{+0.057}_{-0.053}}}
\newcommand{\cOMstat}{\ensuremath{0.272^{+0.018}_{-0.015}}}
\newcommand{\cOLstat}{\ensuremath{0.720^{+0.014}_{-0.014}}}
\newcommand{\cOMsys}{\ensuremath{0.308^{+0.033}_{-0.030}}}
\newcommand{\cOLsys}{\ensuremath{0.693^{+0.024}_{-0.025}}}
\newcommand{\cOMstatf}{\ensuremath{0.278^{+0.017}_{-0.015}}}
\newcommand{\cOLstatf}{\ensuremath{0.717^{+0.012}_{-0.014}}}
\newcommand{\cOMsysf}{\ensuremath{0.306^{+0.026}_{-0.028}}}
\newcommand{\cOLsysf}{\ensuremath{0.694^{+0.019}_{-0.023}}}
\newcommand{\accel}{\ensuremath{99.999}}

\newcommand{\survCalan}{\ensuremath{16}}
\newcommand{\survCFa}{\ensuremath{5}}
\newcommand{\survCFb}{\ensuremath{19}}
\newcommand{\survOther}{\ensuremath{8}}
\newcommand{\survCSP}{\ensuremath{45}}
\newcommand{\survCFc}{\ensuremath{85}}
\newcommand{\survCFd}{\ensuremath{43}}

\newcommand{\colbias}{5}
\newcommand{\wrest}{0.92\pm0.13}

\newcommand{\gps}{\ensuremath{g_{\rm p1}}}
\newcommand{\rps}{\ensuremath{r_{\rm p1}}}
\newcommand{\ips}{\ensuremath{i_{\rm p1}}}
\newcommand{\zps}{\ensuremath{z_{\rm p1}}}
\newcommand{\yps}{\ensuremath{y_{\rm p1}}}
\newcommand{\wps}{\ensuremath{w_{\rm p1}}}
\newcommand{\grizy}{\gps\rps\ips\zps\yps}
\newcommand{\PS}{\protect \hbox {Pan-STARRS1}}

\newcommand{\dLbar}{{\bar d}_{\rm L}}
\newcommand{\dL}{d_{\rm L}}
\newcommand{\zbar}{{\bar z}}
\newcommand{\zSNpec}{z_{\rm SN}^{\rm pec}}

\newcommand{\chisq}{\ensuremath{\chi^2}}
\newcommand{\etal}{et~al.}
\newcommand{\om}{\ensuremath{\Omega_m}}
\newcommand{\ode}{\ensuremath{\Omega_{DE}}}
\newcommand{\oml}{\ensuremath{\Omega_{\Lambda}}}
\newcommand{\kcorr}{\ensuremath{K}-correction}
\newcommand{\kcorrs}{\ensuremath{K}-corrections}
\newcommand{\sg}{\ensuremath{\sigma}}
\newcommand{\si}{\ensuremath{\sigma_{\mathrm{int}}}}
\newcommand{\scriptm}{\ensuremath{\mathcal{M}}}
\newcommand{\scriptmo}{\ensuremath{\mathcal{M}_1}}
\newcommand{\scriptmt}{\ensuremath{\mathcal{M}_2}}
\newcommand{\scriptc}{\ensuremath{\mathcal{C}}}
\newcommand{\gmeg}{\ensuremath{g_M}}
\newcommand{\rmeg}{\ensuremath{r_M}}
\newcommand{\imeg}{\ensuremath{i_M}}
\newcommand{\zmeg}{\ensuremath{z_M}}
\newcommand{\snlsfilts}{\gmeg\rmeg\imeg\zmeg}
\newcommand{\up}{\ensuremath{u^{\prime}}}
\newcommand{\gp}{\ensuremath{g^{\prime}}}
\newcommand{\rp}{\ensuremath{r^{\prime}}}
\newcommand{\ip}{\ensuremath{i^{\prime}}}
\newcommand{\zp}{\ensuremath{z^{\prime}}}
\newcommand{\bd}{BD\ \ensuremath{17^{\circ}\ 4708}}


\newcommand{\xScale}{1.1}   
\newcommand{\xxScale}{1.15}  

\newcommand{\SNANA}{{\tt SNANA}}
\newcommand{\sigMINUS}{\sigma_{-}}
\newcommand{\sigPLUS}{\sigma_{+}}
\newcommand{\xbar}{\bar{x}_1}
\newcommand{\cbar}{\bar{c}}
\newcommand{\SNLS}{SNLS3}
\newcommand{\obss}{observations}
\newcommand{\obs}{observation}

\newcommand{\effspec}{\epsilon_{\rm spec}}
\newcommand{\intonly}{intrinsic-only}

\newcommand{\LCDM}{\mbox{$\Lambda$}CDM}

\newcommand{\ZCUTAVTAU}{0.3}  
\newcommand{\ZMINSYM}{z_{\rm min}}  
\newcommand{\ZMINVAL}{0.02}     
\newcommand{\ZBUBBLE}{0.025}    
\newcommand{\ZCUTLOWZ}{0.02}    


\newcommand{\LOWZMUSIGNIF}{2.4}  
\newcommand{\SDSSMUSIGNIF}{2.5}  

\newcommand{\TMINCUT}{-15}
\newcommand{\TMAXCUT}{+60} 

\newcommand{\sigmutot}{\sigma_{\mu} }
\newcommand{\sigmufit}{\sigma_{\mu}^{\rm fit} }
\newcommand{\sigmuint}{\sigma_{\mu}^{\rm int} }
\newcommand{\sigmudz}{\sigma_{\mu}^{z} }
\newcommand{\sigzspec}{\sigma_{z,spec} }
\newcommand{\sigzpec}{\sigma_{z,pec} }
\newcommand{\sigzflow}{\sigma_{z,flow} }

\newcommand{\sigmurmsVALUE}{0.16}  
\newcommand{\sigzpecVALUE}{0.0012} 
\newcommand{\RMSMU}{{\rm RMS}_{\mu}}

\newcommand{\MUGLOBAL}{\mu_{\rm global}}
\newcommand{\MUSURVEY}{\mu_{\rm survey}}
\newcommand{\MUZBIN}{\mu_{\rm zbin}}
\newcommand{\MUwCDM}{\mu_{{\rm F}w{\rm CDM}}}
\newcommand{\MUZDIF}{\langle\MUZBIN - \MUGLOBAL\rangle}

\newcommand{\hostz}{host-$z$}
\newcommand{\Pfit}{ {\cal P}_{\rm fit} }


\newcommand{\lc}{light curve}
\newcommand{\lcs}{light curves}

\newcommand{\Otot}{\Omega_{\rm tot}}
\newcommand{\OM}{\Omega_{\rm M}}
\newcommand{\OL}{\Omega_{\Lambda}}

\newcommand{\BDFULL}{{\rm BD+17}$^0$4708}
\newcommand{\BD}{{\rm BD+17}}

\newcommand{\SDSS}{SDSS-II}
\newcommand{\mlcs}{{\sc mlcs2k2}}
\newcommand{\mlcsSALTII}{{\sc mlcs-saltii}}
\newcommand{\Mmlcs}{M}

\newcommand{\SALTII}{{\sc salt--ii}}
\newcommand{\chisqmu}{\chi^2_{\mu}}
\newcommand{\dFrestdlam}{\frac{dF_{\rm rest}}{d\lambda}}
\newcommand{\mBstar}{m_B^*}
\newcommand{\SALTIILAMMIN}{2900}
\newcommand{\SALTIILAMMAX}{7000}

\newcommand{\HSTSEARCHFILT}{F850LP\_ACS}

\newcommand{\unc}{uncertainty}
\newcommand{\uncs}{uncertainties}
\newcommand{\Unc}{Uncertainty}
\newcommand{\Uncs}{Uncertainties}


\newcommand{\RVMW}{3.1}


\newcommand{\dw}{\delta w}
\newcommand{\wsystsym}{\sigma_w({\rm syst})}
\newcommand{\wstatsym}{\sigma_w({\rm stat})}
\newcommand{\wtotsym}{\sigma_w({\rm tot})}
\newcommand{\OMsystsym}{\SIGOM({\rm syst})}
\newcommand{\OMstatsym}{\SIGOM({\rm stat})}
\newcommand{\OMtotsym}{\SIGOM({\rm tot})}
\newcommand{\OLsystsym}{\SIGOLAM({\rm syst})}
\newcommand{\OLstatsym}{\SIGOLAM({\rm stat})}
\newcommand{\OLtotsym}{\SIGOLAM({\rm tot})}

\newcommand{\ABAO}{0.469}
\newcommand{\ABAOERR}{0.017}
\newcommand{\BAOCHISQ}{\chi^2_{\rm BAO}}

\newcommand{\RWMAP}{1.710}
\newcommand{\RWMAPERR}{0.019}
\newcommand{\CMBCHISQ}{\chi^2_{\rm CMB}}




\newcommand{\wwwNICMOS}{\tt http://www.stsci.edu/hst /nicmos/documents/handbooks/handbooks/DataHandbookv7}

\newcommand{\wwwACS}{\tt http://www.stsci.edu/hst /acs/documents/handbooks/DataHandbookv4/ACS\_longdhbcover.html}

\newcommand{\wwwMINUIT}{\tt http://wwwasdoc.web.cern.ch/wwwasdoc/minuit/minmain.html}

\newcommand{\wwwSDSS}{\tt http://www.sdss.org/}

\newcommand{\wwwCALSPEC}{\tt http://www.stsci.edu/hst/observatory/cdbs/calspec.html}

\newcommand{\wwwSNANA}{\tt http://www.sdss.org/supernova/SNANA.html}

\newcommand{\wwwTABLES}{\tt http://das.sdss.org/va/SNcosmology/sncosm09\_fits.tar.gz}

%

\newcommand{\dwsym}{\mbox{$\Delta w$}}
\newcommand{\dwsymstat}{\mbox{$\dwsym_{stat}$}}
\newcommand{\dwsymsyst}{\mbox{$\dwsym_{syst}$}}


\newcommand{\dwval}{0.2}      
\newcommand{\dwvalsim}{0.05}  
\newcommand{\dwLOWZSDSS}{0.04}  
\newcommand{\LOWZmudifrms}{0.10}   
\newcommand{\SDSSmudifrms}{0.15}   
\newcommand{\ESSEmudifrms}{0.16}   
\newcommand{\ESSEmudifavg}{0.04}   
\newcommand{\dwSDSS}{0.06}         
\newcommand{\dwSDSSnoU}{0.2}      
\newcommand{\dwmlcsSALTII}{0.2}    
\newcommand{\dwSETg}{0.05}   
\newcommand{\wMLCSg}{-0.84}  
\newcommand{\wSALTg}{-0.89}  


\shorttitle{Systematic Uncertainties of PS1 Sample}
\shortauthors{Scolnic, Rest  et al}
\def\stsci{2}
\def\jhu{1}
\def\cfa{4}
\def\illast{5}
\def\illphy{6}
\def\hawaii{3}
\def\harvard{7}
\def\qub{8}
\def\mpia{9}
\def\waterloo{10}
\def\ucsantacruz{11}
\def\princeton{12}
\def\inaf{13}
\def\lco{14}
\def\ucsb{15}
\def\pitt{16}
\def\durham{17}
\author{
D. Scolnic\altaffilmark{\jhu},
A. Rest\altaffilmark{\stsci},
A. Riess\altaffilmark{\jhu},
M.~E. Huber\altaffilmark{\hawaii},
R.~J. Foley\altaffilmark{\cfa,\illast,\illphy},
D. Brout\altaffilmark{\jhu},
R. Chornock\altaffilmark{\cfa},
G. Narayan\altaffilmark{\harvard},
J.~L. Tonry\altaffilmark{\hawaii},
E. Berger\altaffilmark{\cfa},
A.~M. Soderberg\altaffilmark{\cfa},
C.~W. Stubbs\altaffilmark{\harvard,\cfa},
R.~P. Kirshner\altaffilmark{\cfa,\harvard},
S. Rodney\altaffilmark{\jhu},
S.~J. Smartt\altaffilmark{\qub},
E. Schlafly\altaffilmark{\mpia},
M.~T. Botticella\altaffilmark{\qub},
P. Challis\altaffilmark{\cfa},
I. Czekala\altaffilmark{\cfa},
M. Drout\altaffilmark{\cfa},
M.J. Hudson\altaffilmark{\waterloo},
R. Kotak\altaffilmark{\qub},
C. Leibler\altaffilmark{\ucsantacruz},
R. Lunnan\altaffilmark{\cfa},
G.~H. Marion\altaffilmark{\cfa},
M. McCrum\altaffilmark{\qub},
D. Milisavljevic\altaffilmark{\cfa},
A. Pastorello\altaffilmark{\inaf},
N.~E. Sanders\altaffilmark{\cfa},
K. Smith\altaffilmark{\qub},
E. Stafford\altaffilmark{\jhu},
D. Thilker\altaffilmark{\jhu},
S. Valenti\altaffilmark{\lco,\ucsb},
W.~M. Wood-Vasey\altaffilmark{\pitt},
Z. Zheng\altaffilmark{\jhu},
W.~S. Burgett\altaffilmark{\hawaii},
K.~C. Chambers\altaffilmark{\hawaii},
L. Denneau\altaffilmark{\hawaii},
P.~W. Draper\altaffilmark{\durham},
H. Flewelling\altaffilmark{\hawaii},
K.~W. Hodapp\altaffilmark{\hawaii},
N. Kaiser\altaffilmark{\hawaii},
R.-P. Kudritzki\altaffilmark{\hawaii},
E.~A. Magnier\altaffilmark{\hawaii},
N. Metcalfe\altaffilmark{\durham},
P.~A. Price\altaffilmark{\princeton},
W. Sweeney\altaffilmark{\hawaii},
R. Wainscoat\altaffilmark{\hawaii},
C. Waters\altaffilmark{\hawaii}.
}

\altaffiltext{\jhu}{Department of Physics and Astronomy, Johns Hopkins University, 3400 North Charles Street, Baltimore, MD 21218, USA}
\altaffiltext{\stsci}{Space Telescope Science Institute, 3700 San Martin Drive, Baltimore, MD 21218, USA}
\altaffiltext{\hawaii}{Institute for Astronomy, University of Hawaii, 2680 Woodlawn Drive, Honolulu, HI 96822, USA}
\altaffiltext{\cfa}{Harvard-Smithsonian Center for Astrophysics, 60 Garden Street, Cambridge, MA 02138, USA}
\altaffiltext{\illast}{Astronomy Department, University of Illinois at Urbana-Champaign, 1002 West Green Street, Urbana, IL 61801, USA}
\altaffiltext{\illphy}{Department of Physics, University of Illinois Urbana-Champaign, 1110 W. Green Street, Urbana, IL 61801, USA}
\altaffiltext{\harvard}{Department of Physics, Harvard University, 17 Oxford Street, Cambridge MA 02138}
\altaffiltext{\qub}{Astrophysics Research Centre, School of Mathematics and Physics, Queens University Belfast, Belfast, BT71NN, UK}
\altaffiltext{\mpia}{Max Planck Institute for Astronomy, K\:onigstuhl 17, D-69117 Heidelberg, Germany}
\altaffiltext{\waterloo}{University of Waterloo, 200 University Ave W  Waterloo, ON N2L 3G1, Canada}
\altaffiltext{\ucsantacruz}{Department of Astronomy \& Astrophysics, University of California, Santa Cruz, CA 95060, USA}
\altaffiltext{\princeton}{Department of Astrophysical Sciences, Princeton University, Princeton, NJ 08544, USA}
\altaffiltext{\inaf}{INAF - Osservatorio Astronomico di Padova, Vicolo dell'Osservatorio 5, 35122 Padova, Italy}
\altaffiltext{\lco}{Las Cumbres Observatory Global Telescope Network, Inc., Santa Barbara, CA 93117, USA}
\altaffiltext{\ucsb}{Department of Physics, University of California Santa Barbara, Santa Barbara, CA 93106-9530, USA}
\altaffiltext{\pitt}{PITT PACC, Department of Physics and Astronomy, University of Pittsburgh, Pittsburgh, PA 15260, USA.}
\altaffiltext{\durham}{Department of Physics, University of Durham Science Laboratories, South Road Durham DH1 3LE, UK}

\begin{abstract}
We probe the systematic uncertainties from the \SNIAPSused~Type Ia supernovae (SN\,Ia) in the \PS~(PS1) sample along with \SNIAlowzused~SN\,Ia from a combination of low-redshift surveys.  The companion paper by Rest et al. (2013) describes the photometric measurements and cosmological inferences from the PS1 sample.  The largest systematic uncertainty stems from the photometric calibration of the PS1 and low-z samples.  We increase the sample of observed Calspec standards from 7 to 10 used to define the PS1 calibration system.  The PS1 and \SDSS~calibration systems are compared and discrepancies up to $\sim0.02$ mag are recovered.  We find uncertainties in the proper way to treat intrinsic colors and reddening produce differences in the recovered value of $w$ up to $3\%$.  We estimate masses of host galaxies of PS1 supernovae and detect an insignificant difference in distance residuals of the full sample of $0.037 \pm 0.031$ mag for host galaxies with high and low masses.  Assuming flatness and including systematic uncertainties in our analysis of only SNe measurements, we find $w=$\wSNstasys.  With additional constraints from BAO, CMB (Planck) and $H_0$ measurements, we find $w=\wCsys$~and~$\Omega_m=\fomCsys$ (statistical and systematic errors added in quadrature).  Significance of the inconsistency with $w=-1$ depends on whether we use Planck or WMAP measurements of the CMB: $w_{\textrm{BAO+H0+SN+WMAP}}=-1.124^{+0.083}_{-0.065}$.  
\end{abstract}

\section{Introduction}
\label{sec:intro}
One of the main goals of the \PS~(PS1) Medium Deep survey is to detect and monitor thousands of SN\,Ia in order to measure the equation of state parameter of dark energy, $w=P/\rho c^2$ (where $P$ is pressure and $\rho$ is density).  The first results of this effort are reported in the companion paper by Rest et al. (2013, hereafter R14).  For PS1 and other new surveys to advance our understanding of dark energy, the flood of new SNe must be accompanied by similar improvement in the reduction of systematic uncertainties.
 
Since the initial discovery of cosmic acceleration (\citealp{Riess98}, \citealp{Saul99}), there have been many supernova surveys utilizing multiple passbands and dense time-sampling at both low-z (e.g., CSP,CfA1-4, LOSS, SNFactory\footnote{Carnegie Supernova Project (CSP), Center for Astrophysics (CfA), Lick Observatory Supernova Search (LOSS), Nearby Supernova Factory (NSF)}) and at intermediate and higher-z (e.g., SDSS, ESSENCE, SNLS\footnote{Sloan Digital Sky Survey (SDSS),Equation of State: SupErNovae trace Cosmic Expansion (ESSENCE), SuperNova Legacy Survey (SNLS)}).  While the sample sizes have increased, the systematic uncertainties of these samples now are of nearly equal value to the statistical uncertainties (\citealp{Conley_etal_2011}; hereafter C11).   Nearly all of the systematic uncertainties in the analysis of these samples fall into a small handful of categories: calibration, selection effects, correlated flows, extinction corrections and light curve modeling.   There has been significant recent progress in understanding each of them.  For example, recent studies suggest that properties of host galaxies of SNe appear to be correlated with distance residuals relative to a best fit cosmology (e.g., \citealp{2010ApJ...715..743K}, \citealp{2010MNRAS.406..782S}, \citealp{2010ApJ...722..566L}).  Other studies have shown that supernova colors and brightnesses, long thought to be inconsistent with a Milky Way (MW)-like reddening law, can be explained by a MW-like dust model (\citealp{Folatellietal:2010}, \citealp{Fo11a}, \citealp{Mandel/etal:2011}, \citealp{Chotard_2011}, Scolnic et al. 2013).  

The PS1 Medium Deep Survey has discovered over 1700 SN candidates in its first 1.5 years. Of these, \SNIAPSall ~SNe were spectroscopically identified as Type Ia. Well-sampled multi-band light curves with near-peak observations were measured for~\SNIAPSused~of the spectroscopically confirmed sample.  We include a low-z sample of \SNIAlowzused~SNe to improve our cosmological constraints.  The companion paper by R14 analyzes the photometry of the PS1 light curves, presents the light curve fit parameters and derives constraints on $w$ from a combined data set of PS1 SNe and low-z SNe (hereafter PS1+lz).  In this paper, we augment the work of R14 with a more comprehensive analysis of the systematic uncertainties of $w$.  Values of the matter density $\Omega_m$ and equation-of-state $w$ are recovered with constraints from SNe alone and when we include constraints from measurements of the Cosmic Microwave Background (CMB), Baryon Acoustic Oscillation (BAO) and the Hubble Constant. 

In section \S\ref{sec:overall}, we present an overview of the major systematic uncertainties in our sample, and detail the two approaches towards quantifying these uncertainties.  In section \S\ref{sec:calib_sdss} we analyze the photometric calibration of PS1 and attempt to reconcile the reported calibration discrepancies (\citealp{Tonry12}, hereafter T12) between PS1 and SDSS. We also discuss the data sets in PS1+lz and tension between the various samples.  Accurate simulations of the PS1 survey and expected selection effects for each of the surveys in the combined PS1+lz are given in \S\ref{sec:Malmquist}.  In \S\ref{sec:lcfits} we probe the validity of the two major assumptions of the SALT2 light curve fitter for determining distances to SNe.  In \S\ref{sec:pecvel} we analyze coherent flows of the combined sample for the PS1+lz sample.  Changes to Milky Way extinction maps are presented in \S\ref{sec:extinction}.  Our review and discussion of the dominant uncertainties is given in \S\ref{sec:system} and our conclusions are in \S\ref{sec:conclus}.


\section{Overall Systematics Review}
\label{sec:overall}
\nobreak

\subsection{Data}

The sample analyzed in this paper includes SN\,Ia discovered by PS1 and observed in low-z follow-up programs.  We apply the same selection criteria for the quality and coverage of the light curve observations to these samples as was done in R14.  As detailed in R14, the low-z SN sample is selected from six different samples: Cal\'an/Tololo [\survCalan~SNe] \citep{1996AJ....112.2408H}, CfA1 [\survCFa~SNe] \citep{1999AJ....117..707R}, CfA2 [\survCFb~SNe] \citep{2006AJ....131..527J}, CfA3 [\survCFc~SNe] \citep{2009ApJ...700..331H}, CSP [\survCSP] \citep{2010AJ....139..519C} and CfA4 [\survCFd~SNe] \citep{2009ApJ...700.1097H}.  We also include supernovae not discovered in these surveys but collected as part of the JRK07 \citep{2007ApJ...659..122J} paper [\survOther~SNe].  The PS1 sample contains \SNIAPSused~SNe after selection cuts.  
While the focus of this paper will be on the PS1+low-z sample, we will compare results with the SDSS \citep{Ho08} and SNLS \citep{2010guylightcurves} samples.  For these samples, we apply the same selection criteria from R14.  We make all data used in this analysis publicly available, including light curve fit parameters\footnote{http://ps1sc.org/transients/}. 

External constraints from CMB, BAO and $H_0$ measurements are described in detail in R14.  For all these measurements, we use the Markov chains derived by \cite{Planck13XVI}.  The Planck data set that is quoted includes data from the Planck temperature power spectrum data, Planck temperature data, Planck lensing, and WMAP polarization at low multipoles.  The BAO measurement quoted is from the aggregate of BAO measurements of different surveys, as compiled by \cite{Planck13XVI} and listed in R14.  The $H_0$ measurement is from \cite{Riess1103.2976}.
\subsection{Potential Sources of Systematic Errors}

Here, we briefly enumerate the dominant systematic uncertainties in the PS1+lz sample.  

\textit{Calibration.}  Flux calibration errors are typically the largest source of systematic uncertainty in any supernova sample (C11).  The original PS1 photometric system (T12)  is based on accurate filter measurements obtained in situ.  T12 adjusts the throughput of these measurements on a $<3\%$ scale for better agreement between synthetic and photometric observations of Hubble Space Telescope (HST) Calspec standards \cite{1996AJ....111.1743B}\footnote{http://www.stsci.edu/hst/observatory/cdbs/calspec.html}.  In this paper, we increase the size of the sample of Calspec standards that underpin the HST flux scale from 7 to 10 (adding five, but eliminating two) to reduce the uncertainty in the calibration. 

We call the improved photometric system used throughout this paper the PS1\_14 calibration system.  For the low-z samples, we follow the C11 treatment of photometric systems.  For our total calibration uncertainty, we combine uncertainties from the HST Calspec and Landolt standards, as well as the uncertainties in measurements of the bandpasses and zeropoints.  We also explore noted discrepancies between the PS1 and SDSS photometric systems. 

\textit{Selection Effects.}  Selection effects can bias a magnitude-limited survey, due either to detection limits or selection of the objects for spectroscopic follow-up.  The SNANA simulator\footnote{ SNANA\_v10\_23~@~ \wwwSNANA} \citep{SNANA09} allows us to use actual observing conditions, cadence, and spectroscopic efficiency to mimic our survey.  The spectroscopic efficiency of a survey is particularly difficult to formalize if the survey does not have a single consistent follow-up program that is based on well-defined criteria for selecting targets.

We correct for PS1 selection effects by incorporating the observing history into a simulation and identifying the effective selection criteria that best match the data.  For the low-z sample, we follow the same approach.  For our systematic uncertainty, we explore how well our simulations match the data.

\textit{Light-curve Fitting.}  To optimize the use of SN\,Ia as standard candles to determine distances, most light curve fitters correct the observed peak magnitude of the SN using the width and color of the light curve.  While each fitter accounts in some way for a light curve shape-luminosity relation \citep{phillips93} to correct for the width of the light curve, there is a disagreement between fitters about the best manner to correct for the color of the light curve.  Using SNANA's fitter with the SALT2 model \citep{2010guylightcurves} as the primary light curve fitter, recently Scolnic et al. (hereafter, S13) showed that there is a degeneracy between models of SN color when the intrinsic scatter of SN\,Ia is mostly composed of luminosity variation or color variation.  

The primary method for determining distances uses SALT2 to find light curve parameters and afterwards corrects the distances with the average bias from simulations based on these two models of SN color.  For our systematic uncertainty,  we explore the difference in distances between the two models from when we take the average.

\textit{Host Galaxy Relations}.  Multiple studies have shown relations between various host galaxy properties and Hubble residuals (e.g., \citealp{2010ApJ...715..743K}, \citealp{2010MNRAS.406..782S}, \citealp{2010ApJ...722..566L}).  However, there is no consensus about which host galaxy property is directly linked to luminosity \citep{Ch13}, or whether these correlations may be artifacts of light curve corrections \citep{kimgp}. 

The primary fit does not include any corrections to SN\,Ia distances for host galaxy properties.
For our systematic uncertainty, we explore whether correcting the distances of the SNe in the PS1+lz sample by including information about host galaxies properties is statistically significant.

\textit{MW Extinction Corrections.}  For each SN, we correct for the MW extinction at its specific sky location.  Our primary fit uses values from \cite{1998ApJ...500..525S}, with corrections from \cite{Schlafly11} and the restriction that $E(B-V)<0.5$ in the direction of the SN.  We include systematic uncertainties in the extinction correction from uncertainties in the subtraction of the zodiacal light, temperature corrections, and the non-linearity of extinction corrections \citep{Schlafly11}.

\textit{Coherent Velocity Flows.}  To account for density fluctuations, we correct the redshift of each SN for coherent flows \citep{2004MNRAS.352...61H}. 
The primary fit corrects all redshifts for coherent flows starting at $z_{min}=0.01$.  For our systematic uncertainty, we measure the change in recovered cosmological parameters when we vary the minimum redshift of the sample.

\textit{Other}  Uncertainties not analyzed in this paper, but considered in other studies, include contamination by other types of SN, SN evolution, and gravitational lensing.  Contamination by other types of SN was already discussed in R14, and we apply the same treatment here.  While gravitational lensing should increase the amount of dispersion of the SN\,Ia distances at high-z $\sigma{\mu_{\textrm{lens}}}=0.055z$ \citep{2010MNRAS.405..535J}, selection effects dominate any trend seen at high-z in the PS1 sample.  For SN evolution, this uncertainty is already included in the light curve modeling uncertainty.

Finally, while we include cosmological constraints from the Planck survey \citep{Planck13XVI}, to address systematics in external data sets we compare the results when we include constraints from WMAP \citep{Hinshaw+12}.

\subsection{Error Propagation}

To determine the entire systematic uncertainty of $w$ from the PS1+lz sample, we follow two different approaches towards error accounting.  The first approach follows C11, determining a covariance matrix that includes uncertainties from multiple sources.  The second approach is similar to that of \cite{Riess1103.2976}, which finds the variations of cosmological constraints due to variants in the analysis.  For example, in the second method one may find the different values of $w$ using the PS1 calibration system as stated, or when modified to match the SDSS photometric system.  In the first method, the errors from the PS1 calibration system are propagated.  Since there are a number of discrete choices of how to do steps of the analysis in this paper, we incorporate both methods of error accounting.  Each of these methods is different from the conventional method that adds all the systematic errors in quadrature at the end of the analysis. 

The advantage of the approach shown in C11 is that it properly accounts for covariances between SNe and also for interactions between systematic uncertainties.  The full covariance error matrix is given as:
\begin{equation}
 \mathbf{C} =
  \mathbf{D}_{\mathrm{stat}} +
  \mathbf{C}_{\mathrm{sys}} . \label{eqn:cdef}
\end{equation}
where $\mathbf{D}_{\mathrm{stat}}$ is a diagonal matrix with each element consisting of the square of the intrinsic dispersion of the sample $\sigma_{\textrm{int}}^2$ and the square of the noise error $\sigma_n ^2$ for each SN. $\mathbf{C}_{\mathrm{sys}}$ is the systematic covariance matrix.  C11 further separates  $\mathbf{C}_{\mathrm{sys}}$ into two components, only one of which may be further reduced with more SNe.  For simplicity, we do not separate these components.  Given the Tripp estimator \citep{tripp} and using SALT2 to fit the light curve, 
\begin{equation}
\mu=m_B+\alpha \times x_1 -\beta \times c -M,
\end{equation}
where $m_B$ is the peak brightness of the SN, $x_1$ is the stretch of the light curve, $c$ is the color of the light curve, and $\alpha$, $\beta$ and M are nuisance parameters.  Explanation of the derivations of $\alpha$ and $\beta$ is given in R14.  The systematic covariance, for a vector of distances $\vec{\mu}$, between the i'th and j'th SN is calculated as:
\begin{equation}
 \mathbf{C}_{ij,\mathrm{sys}} = \sum_{k=1}^K
   \left( \frac{ \partial \mu_{i} }{ \partial S_k } \right)
   \left( \frac{ \partial \mu_{j} }{ \partial S_k } \right)
   \left( \sigma_{S_k} \right)^2,
\label{eqn:covar}
\end{equation}
where the sum is over the $K$ systematics $S_k$, $\sigma_{ S_k}$ is the
magnitude of each systematic error, and
$ \partial \mu$ is defined as the difference in distance modulus values after changing one of the systematic parameters.  For example, in order to determine the covariance matrix due to a systematic error of $0.01$ mag in the transmission function of ~\rps~filter, we refit all of the SNe light curves after adding $0.01$ mag to the zeropoint of all observed \rps~values.  Following C11, we do not fix $\alpha$ and $\beta$ when we propagate the systematic covariance matrix.  $\alpha$ and $\beta$ are derived with SALT2mu \citep{Mar11} in the statistical case, though when including the covariance matrix, we write a compatible routine that allows off-diagonal elements.  As given in R14 (Table 4), when attributing the remaining intrinsic scatter to luminosity variation ($\sigma_{\textrm{int}}=0.115$), $\alpha$ and $\beta$ are found to be $0.147\pm0.010$ and $3.13\pm0.12$ respectively.  

Given a vector of distance residuals for the SN sample $\Delta
\vec{\mathbf{\mu}} = \vec{\mu} -
\vec{\mu}(H_0,\Omega_M,\Omega_{\Lambda},w,\vec{z}) $ then $\chi^2$ may be expressed as
\begin{equation}
 \chisq = \Delta \vec{\mu}^T \cdot \mathbf{C}^{-1} \cdot  \Delta \vec{\mu} .
  \label{eqn:chieqn}
\end{equation}
We minimize Eqn. \ref{eqn:chieqn} to determine cosmological parameters that include $H_0, \Omega_M, \Omega_{\Lambda}$ and $w$.  The cosmological parameters are defined in R14 - Eq. 3.  All cosmological parameters quoted in this pair of papers are of the marginalized values and not the minimum $\chi^2$ values.  We assess the impact of each systematic uncertainty by examining the shift it produces in the inferred cosmological parameters.  We also compute the ``relative area" which we define as the area of the contour that encloses 68.3\% of the probability distribution between $w$ and $\Omega_M$ compared to when only including statistical uncertainties.  For this analysis, we assume that the universe is flat.  It is worth clarifying that the relative area may decrease as the contours shift in $w$ vs. $\Omega_M$ space, so relative area alone does not quantify the entire effect of a systematic error.

In the second approach, we redetermine distances based on variations (often binary) in the analysis methods (e.g. \citealp{Riess1103.2976}).  Unlike the method by C11, there is no systematic error component to the error matrix in Eqn.~\ref{eqn:covar}.  Instead, cosmological parameters are found for each difference in analysis approaches.  The overall systematic uncertainty of $w$ from this method is the standard deviation of values for $w$ from the variants to the primary fit.

\section{Systematic Uncertainties in the Absolute Calibration}
\nobreak
\label{sec:calib_sdss}

The flux calibration of PS1 measurements relies on an iterative process that includes work from T12 and \cite{Schlafly12} and is built on by this work and that of R14.  T12 observes several HST Calspec standards \citep{1996AJ....111.1743B} with PS1 and compares the observed magnitudes of these standards to the predicted magnitudes from synthetic photometry.  T12 finds the AB offsets so that the observed magnitudes best matches the synthetic photometry, given tight constraints from measurements on the bandpass edges and shapes.  Catalogs from the fields that contain the Calspec standards are then included as a basis of the relative calibration performed by \cite{Schlafly12},  which uses repeat PS1 observations of stars and solves simultaneously for the system throughput, the atmospheric transparency, and the large-scale detector flat field (called `ubercal').  In this process, new sky catalogs are created not only for the fields in which the Calspec standards are located, but across the entire observable sky including the Medium Deep fields.  

The original observations of Calspec standards by T12 are supplemented by observations of those and other Calspec standards as part of the Pan-STARRs 3Pi survey.  In this work, we determine the AB offsets between the observed magnitudes of the entire set of Calspec standards calibrated by ubercal and the synthetic photometry of these standards.  This iterative process is thus a more accurate test of the absolute flux calibration of PS1.  Once these offsets are found, R14 applies the offsets to the Medium Deep field catalogs, and analyzes further calibration uncertainties that may affect the SN measurements.  Zeropoints for the nightly photometry of the supernovae are determined by comparing the photometry of a single image to the photometry from the Medium Deep field catalog at that location.

The main demarcation between the analysis of Rest et al. and Scolnic et al. when analyzing the calibration uncertainties is that Rest et al. analyzes the uncertainties in the photometric measurements of the stars and supernovae, while Scolnic et al. focus on how these uncertainties propagate to measurements of the absolute calibration of the PS1 system and the supernova distances.

\subsection{Overview of Calibration Uncertainties}

\begin{figure}[h!]
\centering
\epsscale{\xxScale}  
\plotone{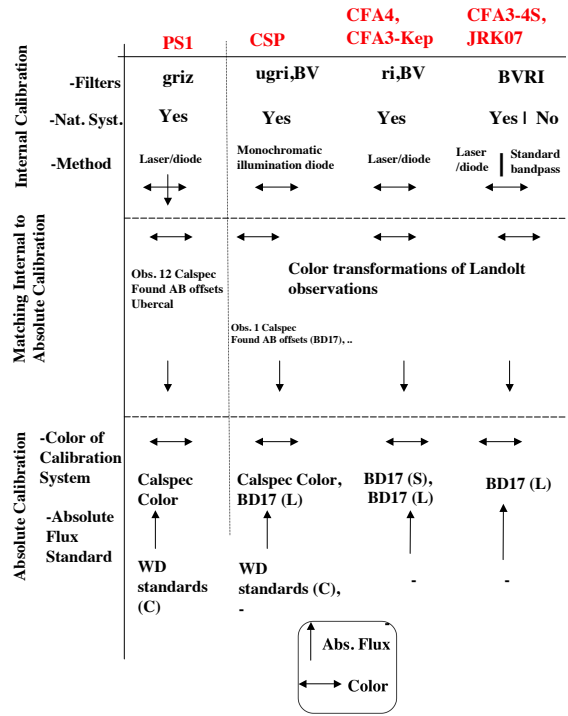}
\caption{A schematic of the calibration of the PS1+lz sample.  The calibration of the various subsamples is broken into three parts: `Internal calibration' (filter measurements), `match between Internal and Absolute calibration', and `Absolute calibration'.  Arrows show whether in each step there may be an uncertainty due to a color measurement or absolute flux measurement.  Directionality of absolute flux arrows show the source of the uncertainty. `C', `L' and `S' represent the HST Calspec, Landolt and Smith standards respectively.}
\label{fig:cal}
\end{figure}

Uncertainties in the calibration of the various samples comprise the largest systematic uncertainty in our analysis.  In Fig.~\ref{fig:cal}, we show a schematic describing the calibration of the various subsamples.  The overall systematic uncertainty in the calibration of our combined PS1+lz sample may be expressed as the combination of three uncertainties.  The first component encompasses systematic uncertainties in the nightly photometry and how well the filter bandpasses are measured.  For the PS1 sample, R14 presents analysis of the systematic uncertainty due to spatial and temporal uncertainties in the nightly photometry.  T12 presents the uncertainty in how well the bandpasses are measured (uncertainty of filter edges $<7~\AA$).  T12 also analyzes uncertainty in the atmospheric attenuation compensation for the filter zeropoints.  

Spatial and temporal variation of the filter bandpasses propagate into our total calibration uncertainty in three ways: how the catalog photometry is determined, how the photometry of the Calspec standards is determined, and how the photometry of the supernovae is determined.  We expect that the uncertainty in the nightly zeropoints to be small.   We find by comparing Pan-STARRs and SDSS photometry that any variation of the PS1 photometry across the focal plane for colors $0.4<g-i<1.5$ is less than 3mmag and is difficult to detect because of noise.  The effect of variation of the filter bandpasses on photometry of the Calspec standards and the supernovae are significantly larger because of the very blue colors of a large fraction of the Calspec standards and the narrow spectral features of these supernovae.  These effects are both considered.

The second major component of the total calibration uncertainty is in determining the flux zeropoints of each filter based on observations of astronomical standards.  Since the accuracy of the internal PS1 measurements of the flux zeropoints is not better than $1\%$, the zeropoints are adjusted so that the observed photometry of HST Calspec standards (e.g., AB - HST Calspec; \citealp{1996AJ....111.1743B}) matches the synthetic photometry of these standards.    Analogously, for the low-z sample, this uncertainty encompasses the accuracy of the color transformation of Landolt standards.  For PS1, the adjustment of the photometry to agree with the synthetic photometry dominates the uncertainty in the filter measurements by T12.  Both uncertainties are included in our analysis.  

The third main component of the total calibration uncertainty is the accuracy in the measurements of the standard stars (e.g., HST Calspec or Landolt).  This is composed of errors in the color of the standard stars and the absolute flux of the standards.  For the PS1 sample and select measurements in the low-z samples, this uncertainty is due to possible errors in measurements of the HST Calspec standards.  For most of the low-z sample, this uncertainty is due to color and absolute flux errors from the realization of the Vega\footnote{The absolute flux of Landolt standards is discussed at the end of the section.} magnitude system as implemented in the standard catalogs of Landolt.  A common flux scale for the PS1 supernovae and low-z SNe can be achieved by the binding between the Landolt catalog and Calspec standards (\citealp{2007AJ....133..768L}).  In a future analysis, we plan to cross-calibrate the Landolt catalog to the PS1 catalog to further improve the flux scales of the different samples.   If we limited our analysis to SNe from a single survey, the overall absolute flux calibration would be degenerate with the absolute peak magnitude of SNe.  But combining distance moduli of SNe from multiple surveys requires that there is a common absolute flux scale.  

R14 presents an error budget for the PS1 photometric system.  Here we explain many of these uncertainties, along with uncertainties of the low-z calibration.  We also detail the derivation of new zeropoint offsets for the PS1 calibration that are used in R14.  The total uncertainty in the recovery of cosmological parameters due to calibration for the PS1+lz sample is given in Table~1.  For each uncertainty described, this uncertainty is independently added to each observed magnitude of the SN, and the light curves are refit.  Afterwards, cosmological constraints are redetermined.  

\subsection{Pan-STARRS Absolute Calibration}
The Pan-STARRS AB magnitude system, as described in T12, is based on small ($<0.03$ mag) adjustments to highly accurate measurements of the PS1 system throughput and filter transmissions measured \textit{in situ}.  Perturbations to each filter transmission are optimized so that `synthetic' photometry, using measurements of filter transmission throughputs and stellar spectra, agrees with observations of HST Calspec standard stars.  To do this, T12 analyzed the PS1 observations of 7 Calspec standards, all observed on the same photometric night.  The error in how well the Calspec SEDs are defined on the AB system as well as the offsets between the observed and synthetic Calspec magnitudes represent the two largest errors in the PS1 calibration system.  T12 finds that the entire systematic uncertainty from absolute calibration in each filter is $\sim0.017$ mag.  We recalculate that value here.

Once the PS1 calibration is defined to be on the AB system, there is an uncertainty from the relative calibration between the fields with Calspec standards to the rest of the sky.  To do the relative calibration, \cite{Schlafly12} use repeat PS1 observations of stars.  Star catalogs created by this process are used in our supernova pipeline.    Internal consistency tests show \cite{Schlafly12} achieve field-to-field relative precision of $<0.01$ mag in \gps,\rps, and \ips~ and $\sim0.01$ mag in \zps~. These errors are included in our overall zeropoint uncertainties, after dividing by $\sqrt{10}$ - the number of fields.  While a following discussion will focus on agreement between the absolute calibration of PS1 and SDSS, it is worth mentioning that \cite{Schlafly12} find greater internal inconsistencies at the $\sim0.01$ mag level in the SDSS photometry than in the PS1 photometry.

\begin{figure}[h!]
\centering
\epsscale{\xxScale}  
\plotone{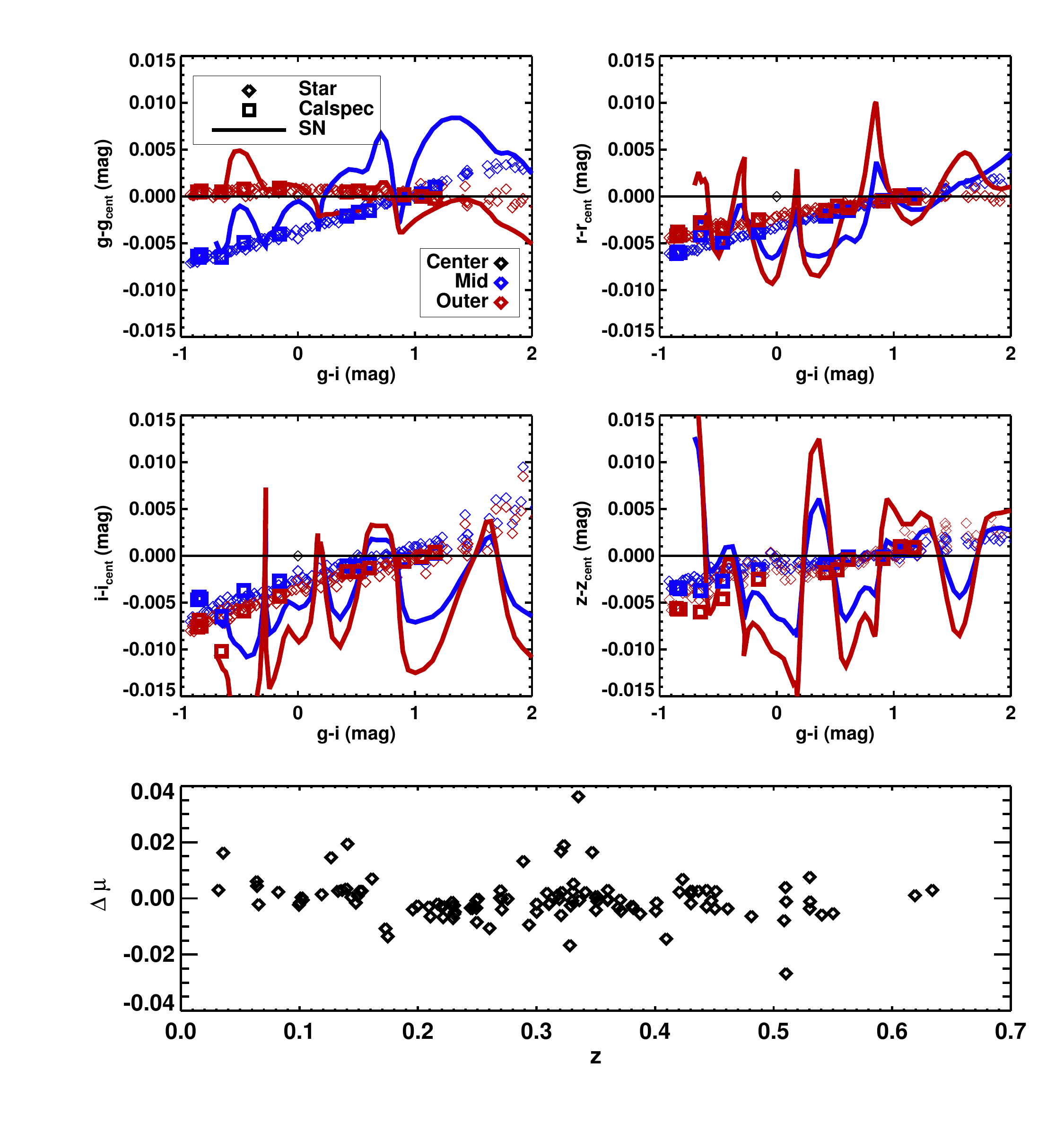}
\caption{(Top panels) The synthetic magnitude differences of Pickles stars, Calspec standards, and a SNIa at different redshifts when the object is observed near the center of the focal plane compared to in an outer annulus.  The smooth change in color of the SNIa is due to red shifting a normal SNIa spectrum.  Filter functions used in this process are from Tonry et al. 2011. (Bottom panel) Changes in distance found for PS1 SNe when the correct filter function at the focal position is used versus the nominal position.}
\label{fig:ps1_focal}
\end{figure}

\begin{figure}[h!]
\centering
\epsscale{\xxScale}  
\plotone{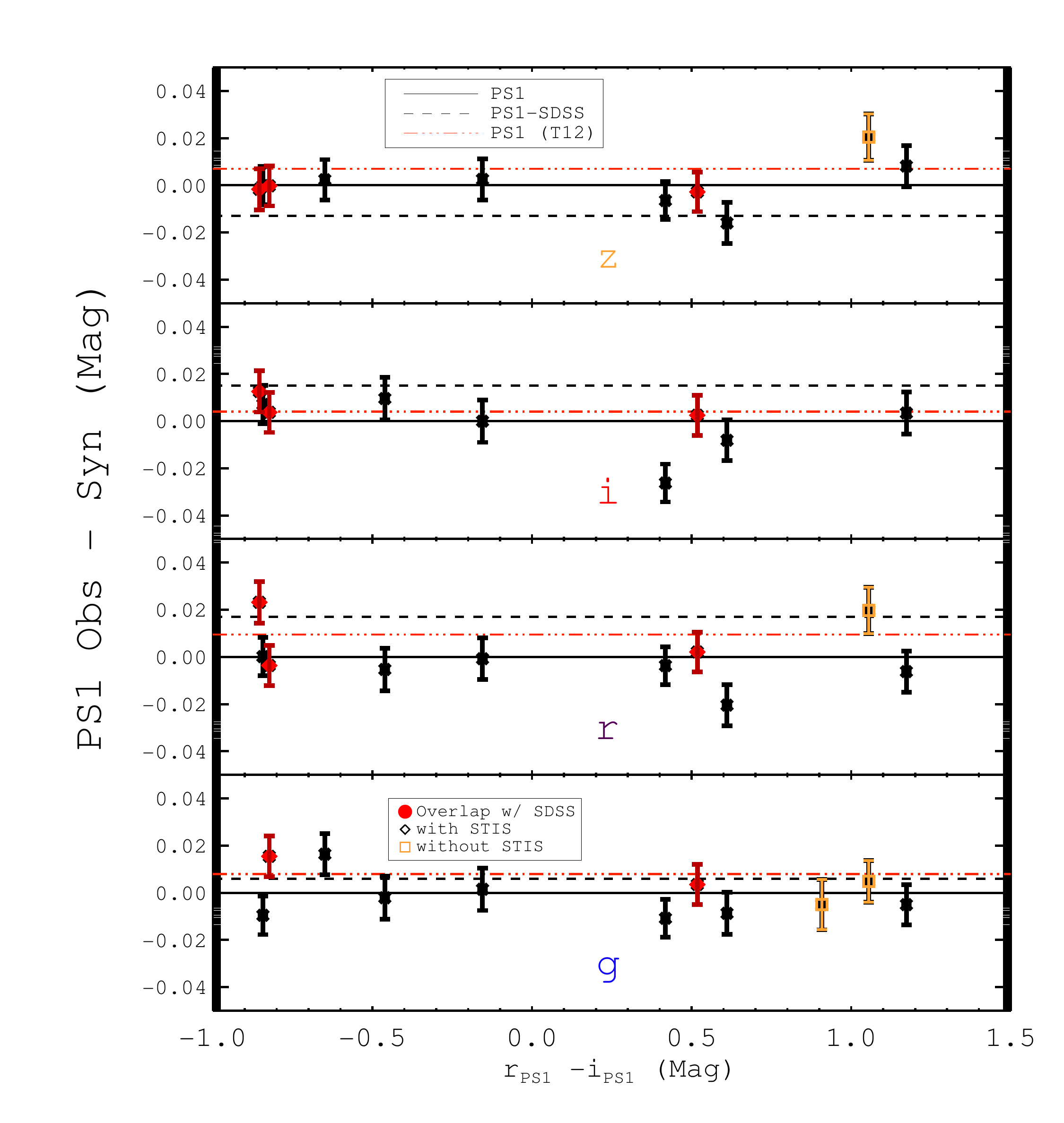}
\caption{The magnitude differences in each passband between observed and synthetic PS1 measurements of 12 Calspec standards.  The solid line represents synthetic photometry from the PS1 photometry, while the dashed line represents AB offsets (given in Table 2) between the SDSS and PS1 absolute calibration.  Standards that are observed by both SDSS and PS1 are shown in red, standards without STIS observed spectra are shown in yellow, and the remaining are shown in black.  AB offsets found in this analysis are such that the discrepancies between the observed and synthetic magnitudes are minimized.}
\label{fig:ps1_cal}
\end{figure}

To improve the PS1 absolute calibration, we analyze a larger sample of Calspec standards that have been observed throughout the PS1 survey.  In total, there are 12 Calpsec standards that have been observed by PS1 in $gri\zps$~that are not saturated in the observations.  These standards were observed so that they avoided the direct center of the focal plane, where there are some unresolved discrepancies as described by R14.  We measure the PS1 magnitudes of the observed Calspec standards in the same way as T12.  We then apply a zeropoint offset obtained by computing the difference of the magnitudes of stars in the fields with the stars in the full-sky star catalogs set by \cite{Schlafly12}.   To avoid a Malmquist bias in the determination of the zeropoint, we empirically determine the magnitude limit at which stars can be used for this comparison.  We also remove any observations of Calspec standards where there is greater than a $0.03$ mag difference between the aperture and PSF photometry, as we found this adequately removes any saturated observations.  As the Calspec standards observed are so bright, we follow T12 and include a $0.005$ mag uncertainty to account for how some of the standards may be near the saturation limit.

To make full use of the observations of Calspec standards, we must consider how the filter functions change across the focal plane.  In Fig.~\ref{fig:ps1_focal}, we show the change in synthetic magnitudes of stars in the Pickles'  library \citep{1998PASP..110..863P}, Calspec standards and supernovae for a given color and position on the focal plane.  We find the variation in synthetic magnitudes for supernovae and Calspec standards may be significantly larger than the variation of stars in the narrow color range used to define the stellar zeropoints.  Therefore, given the measurements of filter functions across the focal plane (measured at $\Delta r=0.15$ deg), we transform the observed magnitudes of the Calspec standards to a uniform system defined at the center of the focal plane.  As shown in Fig.~.~\ref{fig:ps1_focal} , this correction for the blue Calspec standards may be as large as 5 mmag.  This approach is similar to that of  \cite{Betoule2012}. 

For each observation of a Calspec standard, we follow Schlafly et al. and assign an observational error of $0.015 + \sigma_{mag-psf}$, which Schlafly et al. finds adequately describes the scatter seen in their star catalogs.  To determine the net adjustment needed for each passband, we find the weighted difference of the observed and synthetic magnitudes of the Calspec standards.  For this process, we add an additional uncertainty of $0.008$ mag to each difference in order to represent the uncertainty in the ubercal process.  In Appendix A, we present the entire set of Calspec standards observed and their synthetic magnitudes in the PS1 system\footnote{PS1 passbands can be found on the ApJ webpage for T12.}, as well as the observed magnitudes.  From Figure \ref{fig:ps1_cal}, we find that corrections should be added to the zeropoints of observations in each filter (given by T12 and  \citealp{Schlafly12}) such that $\Delta g_{PS1} = -0.008$, $\Delta r_{PS1} = -0.0095$, $\Delta i_{PS1} = -0.004$, $\Delta z_{PS1} = -0.007$.  These adjustments represent the weighted difference of the observed and synthetic magnitudes of the Calspec standards.  R14 includes these offsets in their light curve fits; we call the new calibration `$\textrm{PS1}_{\textrm{13}}$'.  We determine the uncertainties in the mean for all four passbands to be [0.0085,0.0050,0.0060,0.0025] mag.  These uncertainties are included in the overall calibration uncertainty table of R14 (T).

T12 finds consistency of $\sim0.01$ mag among their 7 Calspec stars used to define the AB system.   For two of the standards, 1740346 and  P177D, T12 noticed disagreement at the 0.02 mag level in \ips.  T12 explained that this disagreement may be due to the discontinuity at 800 nm where the STIS spectra gives way to NICMOS in the Calspec SEDs.  With our larger sample of Calspec standards, we can see that the disagreement T12 noticed is most likely not due to 1740346 and P177D, but rather the three `KF' stars (KF08T3, KF01T5, KF06T2), which are red stars with $r-i\sim0.3$.  In the analysis done by T12, the spectra of the KF stars did not include STIS data, which covers the optical spectrum.  We find that with an updated STIS spectrum of KF06T2, the synthetic photometry is corrected by $\sim0.02$ mag, in better agreement with observations.  While we present the entire set of HST Calspec standards observed, we exclude the two KF stars without STIS measurements from our absolute calibration.

Including the uncertainties described above, R14 finds the combined uncertainty for each filter in quadrature is $\sim 0.012$ mag.  Uncertainty in $\gps$ appears to have the largest effect on the cosmological constraints compared to the other passbands.  Interestingly, a calibration error in \rps~appears to have a different effect than the other passbands because the change in distance due to peak brightness in this filter cancels out the change in distance due to color (for $>50\%$ of redshift range).  The effect on recovered cosmological parameters from $gri\zps$ together (labelled `PS1 ZP +Bandpasses' in Table 1) increase the relative area of the constraints by $40\%$ (SN only).

We also consider the effects of the third component of the total systematic uncertainty in calibration (bottom level of Fig. 1): that of the calibration of the HST Calspec standards to the AB system.  T12 states this uncertainty is $0.013$ mag for all filters.  We find a more appropriate solution is to take the uncertainty as the inconsistency between the synthetic photometry of the STIS measurement of BD17 and observed photometry from ACS given in \cite{2004AJ....128.3053B}.  This error is explained by the STIS flux for BD17 that continuously drops from a multiplicative factor times the flux of $1.005$ at $4000~\AA$~to $0.985$ at $9500~\AA$\footnote{\cite{2004AJ....128.3053B} argue that this error may be partly composed of errors from ACS bandpasses, so our systematic uncertainty here is likely conservative.}. This is similar to the 0.5\% slope uncertainty stated by \cite{BohlinHartig}.  Additionally, there is a measurement error in the repeatability of the individual measurements with STIS spectra, on the order of $0.005$ mag \citep{Betoule2012}.   The impact of these uncertainties of the Calspec standards is given in Table 1 and increase the relative area by $\sim5\%$. Finally, the absolute flux of the Calspec system itself must be taken into account.  This uncertainty will be considered as part of the low-z discussion later in this section (given as `Abs. ZP' in Table~1). 

\begin{deluxetable*}{l|ccc|ccc}
\tablecaption{Calibration Systematics
\label{tab:calcheck2}}
\tablehead{
\colhead{Systematic} &
\colhead{$\Delta \Omega_M$} &
\colhead{$\Delta w$ } &
\colhead{Rel. area }  &
\colhead{$\Delta \Omega_M$ } &
\colhead{$\Delta w$} &
\colhead{Rel. area}  \\
~ & ~ & SN Only & ~ & ~ & SN+BAO+CMB+$H_0$ & ~ 
}
\startdata
Stat. Only & 0.000 & 0.000 & 1.000 & 0.000 & 0.000 & 1.000 \\
 & $(\Omega_M=0.223^{+0.209}_{-0.221})$ & $(w=-1.010^{+0.360}_{-0.206})$ & ~ & $(\Omega_M=0.284^{+0.010}_{-0.010})$  & ($w=-1.131^{+0.049}_{-0.049}$) & \\
\\
\hline
\\
PS1 ZP+Bandpass & 0.005 & -0.038 & 1.285 & -0.003 & -0.025 & 1.221 \\
PS1 g & 0.008 & -0.038 & 1.074 & -0.001 & -0.013 & 1.081 \\
PS1 r & 0.004 & -0.006 & 1.028 & 0.001 & 0.001 & 1.000 \\
PS1 i & 0.000 & 0.002 & 1.119 & 0.000 & -0.005 & 1.079 \\
PS1 z & -0.004 & -0.001 & 1.068 & -0.001 & -0.009 & 1.080 \\
Low-z ZP+Bandpass & -0.001 & -0.012 & 1.070 & -0.001 & -0.010 & 1.085 \\
Landolt Color & -0.000 & 0.004 & 1.013 & 0.001 & 0.001 & 1.004 \\
Calspec Uncertainty & -0.001 & -0.025 & 1.089 & -0.002 & -0.020 & 1.126 \\
Abs. ZP & -0.000 & 0.004 & 1.004 & 0.001 & 0.001 & 1.001 \\
SALT2 calibration & 0.029 & -0.063 & 1.202 & 0.000 & -0.002 & 1.033 \\
\\
\hline
\\
All Cal. systematics & 0.024 & -0.093 & 1.566 & -0.004 & -0.035 & 1.309 \\
 & $(\Omega_M=0.248^{+0.210}_{-0.165})$ & $(w=-1.105^{+0.435}_{-0.305})$ & ~ & $(\Omega_M=0.280^{+0.013}_{-0.012})$  & ($w=-1.166^{+0.067}_{-0.069}$) & \\

\enddata
\tablecomments{ Individual systematic uncertainties for each of the PS1 passbands as well as the systematic uncertainties for each low-z sample.  $\textrm{RelativeArea}$ is the size of the contour that encloses 68.3\% of the probability distribution between $w$ and $\Omega_M$ compared with that of statistical-only uncertainties.}
\end{deluxetable*}

\subsection{Absolute Calibration Agreement Between PS1 and SDSS}

 Surveys like SDSS, CSP and SNLS have recently undertaken large, collaborative efforts (\citealp{2012AJ....144...17M}, \citealp{Betoule2012}) to improve the agreement between their respective calibration systems.  Here we focus on the consistency of the absolute calibration between PS1/SDSS as the absolute calibration differences between SDSS/SNLS \citep{Betoule2012} and SDSS/CSP \citep{2012AJ....144...17M} have been shown to be less than $1\%$.  As SDSS photometry has been defined to be on the AB system, this analysis is an alternate diagnostic to quantify the accuracy of the PS1 photometric system itself.
 
T12 compares the Pan-STARRS1 magnitudes of stars in the MD09 field with those tabulated by SDSS as part of Stripe 82.  They note $\sim0.02$ mag offsets at 3-4$\sigma$ after transforming the SDSS DR8 catalogs into the PS1 system with linear terms in color\footnote{For color transformations: PS1 filter transmissions from T12, SDSS filter transmissions from \cite{2010AJ....139.1628D}~and Pickles star spectra \citep{1998PASP..110..863P}}. T12 conclude that discrepancies are most likely due to errors within the SDSS calibration system.  
\begin{figure}[h]
\centering
\epsscale{\xxScale}  
\plotone{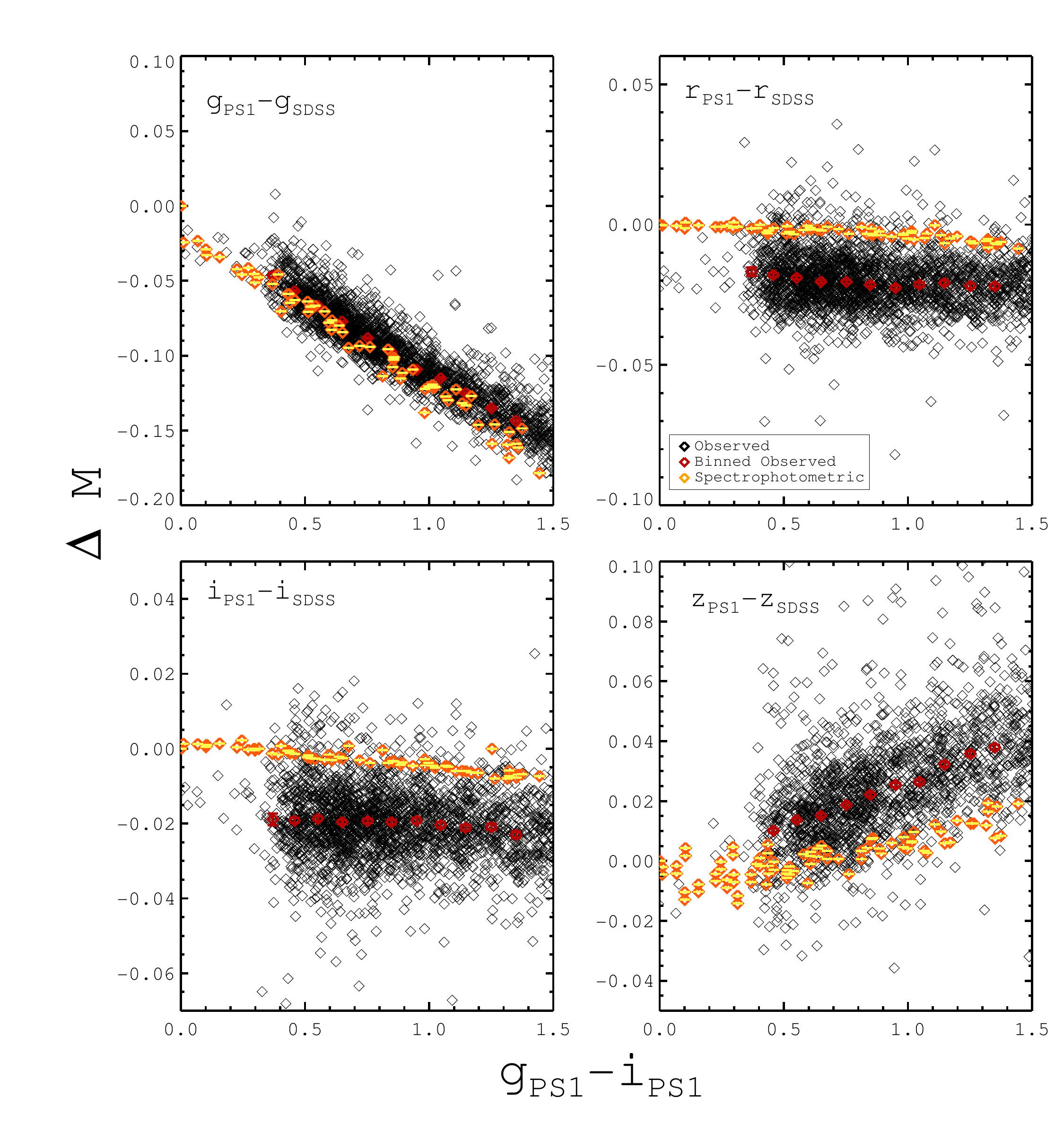}
\caption{Passband magnitude differences between PS1 catalog (T12 catalogs with ubercal zeropoints) and the most up-to-date SDSS S82 Catalog (from \citealp{ivezi_sloan_2007}, \citealp{Betoule2012}) are shown in yellow.  In yellow are the synthetic spectrophotometry differences for a set of Pickles \citep{1998PASP..110..863P} stars using the T12 PS1 passbands and \cite{2010AJ....139.1628D} passbands. }
\label{fig:ps1_sdss1}
\end{figure}
For comparisons between SDSS and PS1, we repeat the analysis in T12, now using the most up to date S82 catalogs from \cite{ivezi_sloan_2007} with AB offsets from \cite{Betoule2012}.  Discrepancies between these two systems are shown in Fig.~\ref{fig:ps1_sdss1}.  The offsets between the calibration zeropoints of PS1 and SDSS are given in Table~\ref{tab:magtest2} and are up to $\sim0.02$ mag in \rps.  \cite{Betoule2012} redefines the SDSS AB system using SDSS PT observations of 7 Calspec standards.  While T12~explains that the absolute calibration of SDSS DR8 may be biased from using SDSS SEDs, \cite{Betoule2012} uses HST Calspec spectra to define the flux system so there should not be an issue. 
\begin{deluxetable}{lcc}
\tablecaption{PS1 Photometric Consistency Checks 
\label{tab:magtest2}}
\tablehead{
\colhead{Filter} &
\colhead{From T12} &
\colhead{$\textrm{PS1}_{13}$ +B12} \\
~ & [Mag] & [Mag]  \\
}
\startdata
  \gps &  \phantom{-}0.014 &\phantom{-} 0.0095 \\
  \rps & $-$0.019 & $-$0.017 \\
  \ips &    \phantom{-}0.008 & -0.015 \\
  \zps &    \phantom{-}0.015 & \phantom{-}0.016  \\
\enddata
\tablecomments{AB offsets from comparisons of PS1 and color-transformed SDSS and  catalogs.  The first column shows the offsets obtained in T12, the second column shows comparisons between the $\textrm{PS1}_{13}$ photometry SDSS S82 photometry as released in  \cite{Betoule2012}.  These offsets are given in the form $m_{\textrm{PS-obs.}}-m_{\textrm{SDSS-obs.}}-(m_{\textrm{PS-syn.}}-m_{\textrm{SDSS-syn.}})$ where $m$ is the magnitude in any given filter.}
\end{deluxetable}

To further probe the inconsistency between the PS1 and SDSS calibration systems, in Appendix A, we compare synthetic and observed magnitudes for the SDSS standards in the same way we did for PS1.  In both Fig.~\ref{fig:ps1_cal} and Appendix A, we show how, for PS1 and SDSS, the zeropoints should be shifted into agreement with the color-transformed system of the other.  We note that for SDSS, the dispersion of the differences between synthetic and observed photometry is smaller than that for PS1, and likely does not explain the difference in absolute zeropoints.  In the comparison of PS1 and SDSS catalogs shown in Fig.~\ref{fig:ps1_sdss1}, the differences have a very small dependence on the color $g-r$ ($<5$ mmag for $g-r < 1.2$, highest in the $z$ band).  This result is encouraging that while the absolute zeropoints of the filters are in disagreement, the filter transmission curves appear to be well measured for both systems.  Also, the zeropoint discrepancies do not appear to be correlated across filters.

There are three Calspec standards observed by both PS1 and SDSS: P177D, GD71 and GD153.    The discrepancies between the PS1 and SDSS observations of these standards are very similar to the overall discrepancies in the calibration of these two systems and do not provide enough leverage to understand the source of the differences.  Therefore, more work must be done to understand the disagreement between the PS1 and SDSS calibration.  One possible cause may be due to non-linearities with these particular observations of very bright standards, which T12 estimated for the PS1 observations to be up to $\sim0.005$ mag.  We conclude that when combining data from the PS1 and SDSS surveys that the AB offsets between the two must be taken into account.  We find that the change in $w$ when the PS1 calibration system is chosen to be in agreement with SDSS is $\Delta w = +0.018$ with constraints only from SN measurements, and $\Delta w =-0.006$ when including CMB, BAO and $H_0$ constraints (due to how constraints combine).  This difference for the SN-only constraints is the largest variant in our analysis. 

\subsection{Nearby Supernova Sample Absolute Calibration}
\nobreak
\label{sec:datasets}

We rely on analysis of past studies, in particular C11 and \cite{2009ApJS..185...32K}, for our understanding of the calibration systematics of low-z surveys. We discuss our additions to the growing low-z sample: the CfA4 survey, a recalibration of the CfA3 survey, and a larger set of CSP SNe.  Given the U-Band systematic error discussed in \cite{2009ApJS..185...32K}, we follow the C11 decision to not use rest-frame observations in the U-band.

Each of the newly added nearby samples has photometry on its natural system.  For each sample, the absolute calibration is defined by the magnitudes of the fundamental flux standard BD17.  For CSP, the magnitudes of BD17 are given in the natural system.  For CfA3 and CfA4,  we use the linear transformations from the Landolt \citep{1992AJ....104..340L} and Smith \citep{smith02} colors to the natural system to determine the magnitude of BD $17^{\circ} 4708$.  These transformations are given in \cite{2009ApJ...700..331H}.  The magnitudes of BD17 given in \citet{1992AJ....104..340L} are transformed to calibrate the \emph{BV} bands, and the magnitudes of BD17 given in \citet{smith02} are transformed to calibrate the \emph{r'i'} bands. 

We note two peculiarities with the CfA3 and CfA4 samples.  To analyze these two samples, we take passbands defined by \citet{cramer12} in which highly precise measurements are obtained of the telescope-plus-detector throughput by direct monochromatic illumination.  This method, like that done in T12 is based on \citet{stubbs06}. However, the CfA4 survey must be broken into two separate time periods because it was found that a warming of the CCDs of KeplerCam to remove contamination in May 2011 `produced a dramatic difference' in the response function of the camera \citep{2009ApJ...700.1097H}.  This difference is quantified by measurements of the $V$, $V-i'$, $U-B$ and $u'-B$ color coefficients between the Landolt/Smith measurements and the natural system.  Therefore, we use a set of transmission functions for before August 2009 and after May 2011, when the system was measured to be consistent, and a separate set of transmission functions between these two dates (Hicken et al. 2012).  The second peculiarity is that when analyzing the CfA4 light curves, we found the uncertainties for each observation to be on average roughly $\sqrt {3}$ larger than that of CfA3, a surprising result considering the similarity of the surveys.  We discovered this was due to a change in uncertainty accounting in the software pipeline based on the number of image subtractions done for each observation (Hicken, private communication).  We have returned the uncertainty propagation method to that used for CfA3, which we believe to be correct.   

For our uncertainties in the low-z bandpasses and zeropoints (top level of Fig.~\ref{fig:cal}), we follow the analysis of C11.  We use the shifted Bessell bandpasses found empirically by C11 with uncertainties of 12~\AA~(edge locations) for the JRK07 sample and adopt zeropoint uncertainties of $0.015$ mag.  For CSP, the uncertainties in the bandpasses are taken from \cite{2010AJ....139..519C}  and the uncertainties in the zeropoints for each filter are $0.008$ mag, as given in C11.  While more work must be done to better determine this zeropoint uncertainty, this result is consistent with the small discrepancies of $\sim0.01$ mag seen between the CSP and SDSS samples \citep{2012AJ....144...17M}.  We take the zeropoint uncertainties for the CfA4 sample given in Hicken et al. 2011 of 0.014, 0.010, 0.012, 0.014, 0.046 mag in $BVr'i'u'$, which are larger than those found by C11 for the CfA3 sample of $0.011,0.007,0.007,0.007$ mag for $BVRr'$.  Since the uncertainty of the CfA4 bandpasses measured by Cramer et al. (in prep) has not yet been given, we fix this uncertainty to be that found by T12 for the PS1 passbands (3~\AA), as Cramer et al. and T12 perform very similar measurements to determine the instrument response.

While the absolute flux of the HST Calspec standards is defined by the AB system, the absolute flux of the Landolt standards is not well defined.  Although the Landolt measurements are self-consistent, it is not known exactly how the absolute flux was defined.  Therefore, there may be discrepancies between the absolute flux of these different sets of standards (see bottom level of Fig.~\ref{fig:cal}). We follow the analysis of \cite{2007AJ....133..768L} of the calibration agreement between the Landolt catalog and HST observations of Calspec standards for an uncertainty of $0.006$ mag between the absolute fluxes of the two samples.  For the difference between the Smith and AB systems, we take the uncertainty in determining the AB offsets for the SDSS sample of $\sim0.004$ mag \citep{Betoule2012}.  We also account for uncertainties in the colors of Landolt measurement of BD17 itself of $\sim0.002$ mag (Regnault et al. 2009).  This last uncertainty could be reduced by defining the low-z samples using more standards besides BD17 (a subdwarf star), which will be done in a future work.  
\subsection{Further calibration systematics and impact}
While we have discussed the entirety of calibration errors that affect the measurements of the supernova in our sample, we must also propagate how calibration errors affect the SALT2 light curve model that we use to fit distances.  To do so, we refit our entire SN sample with 100 variants of the SALT2 model based on the calibration errors of the training sample used to determine the model (Guy10).  These variants were provided by the SALT2 team.  For the total systematic from the SALT2 calibration error, we sum the covariance matrices over all of the iterations and then divide by the total number of iterations.  This impact of this uncertainty is quite large with respect to our other calibration uncertainties as it increases our $w$ versus $\Omega_M$ constraints for the SN-only case by $>15\%$.  

\begin{figure}[h!]
\centering
\epsscale{\xxScale}  
\plotone{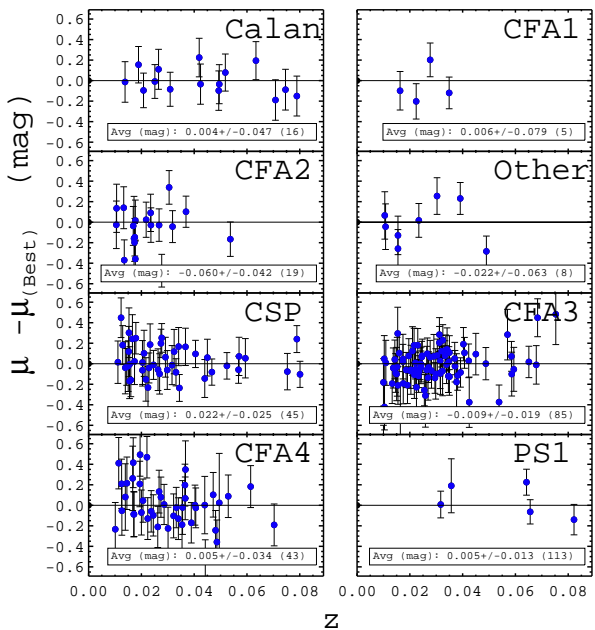}
\caption{The Hubble Residuals for each supernova sample at low-z.  On each panel, the tension between this set and the others is shown.}
\label{fig:lowz_tension}
\end{figure}

The impact on the recovery of cosmological parameters from all of the uncertainties discussed above are presented in Table~1.  Uncertainties in the low-z transmission measurements are significant (SN. only relative area $\sim1.07$), though do not have as great an increase on the relative area as the uncertainty in the PS1 transmission throughputs.  We present the distance residuals from the best fit cosmology for each low-z survey in Fig.~\ref{fig:lowz_tension}.  We refer here to R14 (section 7.2) which details the quality culls on the light curves and reduces the number of light curves significantly.  The intrinsic dispersion ($\sigma_{\textrm{int}}$) , RMS and effects on retrieved cosmology from removing a particular subsample are all shown in Table~\ref{tab:lowzrem}.  We note that the $\sigma_{\textrm{int}}$  of the PS1 sample ($\sigma_{\textrm{int}}=0.07$) is lower than in other samples, though is closest to the CSP sample. We also note that the CfA4 sample appears to have a larger scatter ($\sigma_{\textrm{int}}=0.22$) than the other samples.  The various values of $\sigma_{\textrm{int}}$ may be due to over or under-estimation of calibration uncertainties. Part of this trend may also be due to a low-z Malmquist bias, which will be discussed in a later section. The maximum tension between the low-z subsamples is about $<2\sigma$ from the mean. REF27: There are 23 SNe observed by both CSP and CfA3, and the mean difference in distances for these SNe is $0.026\pm0.03$ mag (CSP-CfA3) with an RMS of 0.16 mag.  We only allow a single distance for a given supernova, and choose based on which has better cadence near peak.  Following C11, we add different $\sigma_{\textrm{int}}$ values to the photometric uncertainty of the SN distances for our high and low-z samples, though not for the individual low-z subsamples.

\begin{deluxetable*}{lllllll}
\tablecaption{Effects of Removing Low-z Sample on Cosmology
\label{tab:lowzrem}}
\tablehead{
\colhead{Sample:} &
\colhead{$\sigma_{\textrm{int}}$} &
\colhead{$\textrm{RMS}$} &
\colhead{$\Delta \Omega_m$} &
\colhead{$\Delta w$} &
\colhead{$\Delta \Omega_m$} &
\colhead{$\Delta w$} \\
~&~&~&(SNe only)&(` ')&(SN+H0+CMB+BAO )&(` ')\\
}
\startdata
JRK07 & 0.125 & 0.180 & -0.008 & 0.043 & 0.004 & 0.022 \\
CSP & 0.105 & 0.149 & 0.004 & -0.038 & -0.003 & -0.021 \\
CFA3 & 0.115 & 0.185 & -0.009 & 0.016 & 0.000 & 0.000 \\
CFA4 & 0.170 & 0.218 & 0.009 & -0.030 & -0.001 & -0.008 \\
PS1 & 0.070 & 0.179 & - & - \\

\enddata
\tablecomments{For the PS1+lz sample, we show $\Delta \Omega_m$ and $\Delta w$ (SN constraints as well as full SN+H0+CMB+BAO constraints) when we remove one of the subsamples, and keep the rest of the sample intact.  We also give the intrinsic dispersion of each subsample and the RMS.}
\end{deluxetable*}

\section{Selection Effects}
\nobreak 
\label{sec:Malmquist}

\subsection{PS1 Selection Bias}
To make use of SNe discovered near the magnitude limits of the PS1 supernova survey, we account for the selection bias towards brighter SNe.  These SNe may be brighter, even after the light-curve shape and color corrections, due to the intrinsic dispersion of SN\,Ia and/or noise fluctuations.  To determine the correction for this selection bias, we simulate the PS1 Medium Deep Survey and the spectroscopic follow-up.  We apply the \SNANA\ Monte-Carlo (MC) code to generate realistic SN\,Ia light curves with the same cadence, observing conditions, non-Gaussian PSFs (R14), and zeropoints as our actual data.  All simulations are based on a standard $\LCDM$ cosmology with $w=-1$, $\OM=0.3$, $\OL=0.7$, $H_0=70$ and an absolute brightness of a fiducial SN\,Ia of $M_B=-19.36$ mag \cite{SNANA09}.  Varying these initial conditions has negligible impact on our derived Malmquist bias.  To simulate the intrinsic scatter of SN\,Ia, we use the SALT2 model (hereafter called Guy10, \citealp{2010guylightcurves}), which will be discussed in detail in the next section. 

While the photometric selection bias may be inferred from survey conditions, the spectroscopic selection bias must be estimated with an external model for selection efficiency.  We first find the minimum Signal-to-Noise ratio (SNR) values of all PS1 SNe in the photometric and spectroscopic samples. For photometric detection of a SN, we require that there are $2$ measurements in any filter with $S/N > 5$.  For spectroscopic follow-up, we find from our sample that there are always at least $2$ measurements in any filter with $S/N > 7$ and $1$ measurement in any filter with $S/N > 9$.  The spectroscopic efficiency function, $E_{\textrm{spec}}$, is applied after the SNR cuts are in place and is split into two  parts because there were separate PS1 follow-up programs for high and low-z SNe:
\begin{multline}
E_{\textrm{spec}} (z\le0.1)=e^{-z/z1} ; \\
E_{\textrm{spec}} (z>0.1)=1.0/[1 + (\rps-19.0)]^{r1}.
\label{eqn:sel}
\end{multline}
Here, $z$ is the redshift and $\rps$ is the observed peak magnitude.  We vary $z1$ and $r1$ to optimize agreement between the number of observed SNe with redshift for the data and MC.  We find that $z_1=0.05$ and $r_1=2.05$.  

\begin{figure}[h!]
\centering
\epsscale{\xxScale}  
\plotone{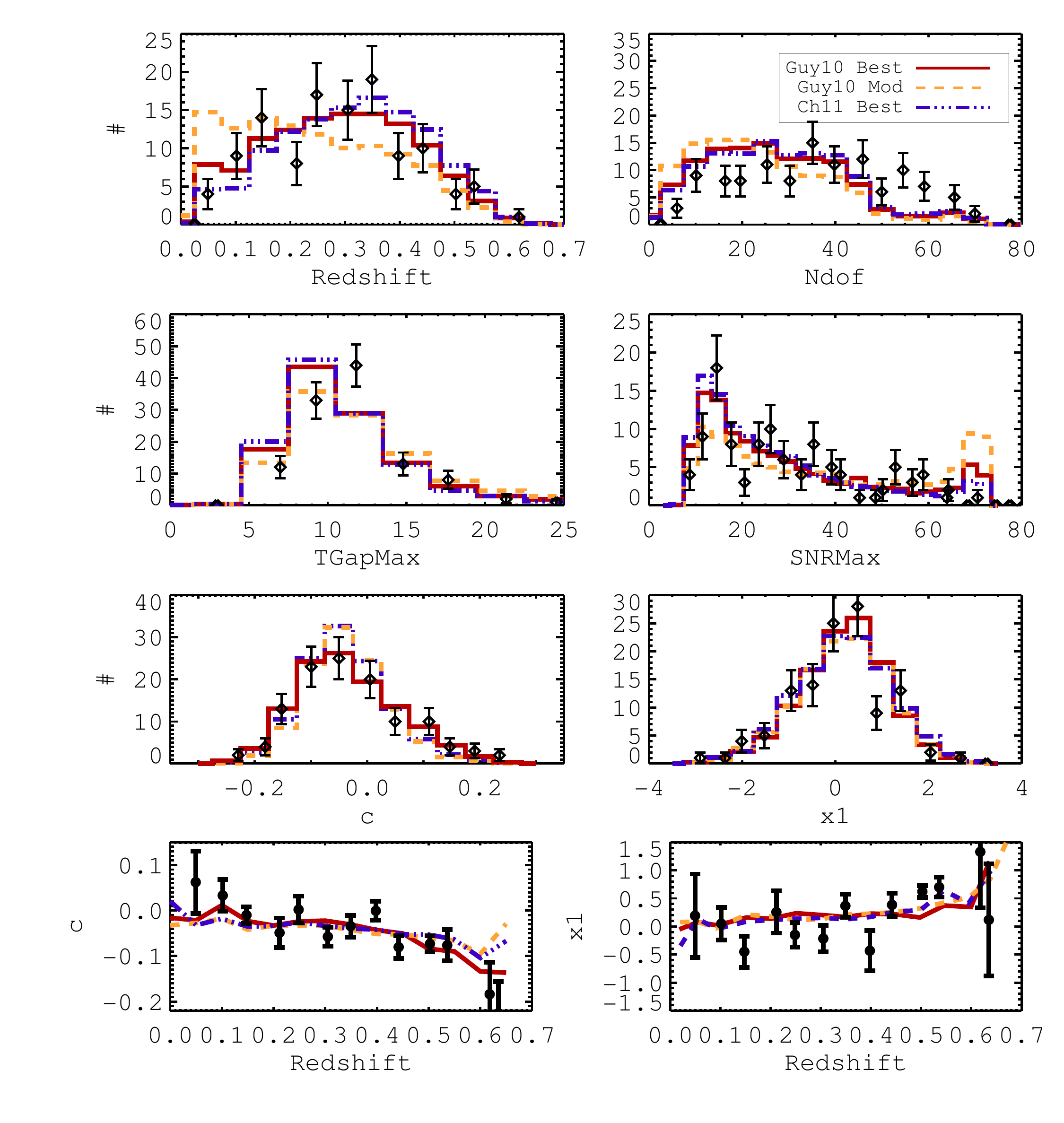}
\caption{ Comparison of distributions for PS1 data (points) and 
    MC (histogram), where each MC distribution is scaled to 
    have the same sample size as the data.  We show the simulation of the survey assuming a Guy10 scatter model for the intrinsic dispersion(red), the $2\sigma$ difference for adjusted selection efficiency (orange), and assuming a Chotard11 scatter model (blue).
    The distributions are 
    redshift, 
    number of degrees of freedom in the light curve fit, 
    maximum rest-frame time difference (gap) between {\obss},
    maximum $S/N$, 
    fitted \SALTII\ color ($c$) and light-curve shape parameter ($x_1$).
    The bottom two panels show the
    \SALTII\ color ($c$) and shape parameter ($x_1$) versus redshift. }
\label{fig:simverify}
\end{figure}

A comparison of properties of the SNe recovered (e.g. redshift distribution, SNR distribution) from the PS1 simulation to the actual data is shown in Fig.~\ref{fig:simverify}.  Overall we find very good agreement between our data and MC.  Discrepancies in these comparisons, especially that of cadence, may be explained by observational effects like masking and saturation which we do not model in the simulation.  To determine the systematic uncertainty of the selection bias, we vary the spectroscopic selection function at high-z and find where the data/MC comparison of the redshift distributions is worse than our best fit by $2\sigma$.  We conservatively choose $2\sigma$ here because our spectroscopic follow-up program was done with multiple telescopes and assuming one follow-up program may lead to a bias.  For this test, we exclude $z<0.25$ as the PS1 selection bias at low-z should be negligible, and the low-z sample dominates at $z<0.1$.  Results of this simulation ($r_1=2.5$ in Eqn.~\ref{eqn:sel}) are over-plotted in Fig.~\ref{fig:simverify}.  We fix the uncertainty at a given redshift in the selection bias to be the difference between the selection bias for these two simulations.  The Malmquist bias is shown in the following section.  The uncertainty in the selection bias is included in the overall uncertainty budget at the end of the analysis in Table~\ref{tab:dom}.  

\subsection{Low-z Selection Effects}
\begin{figure}[h!]
\centering
\epsscale{\xxScale}  
\plotone{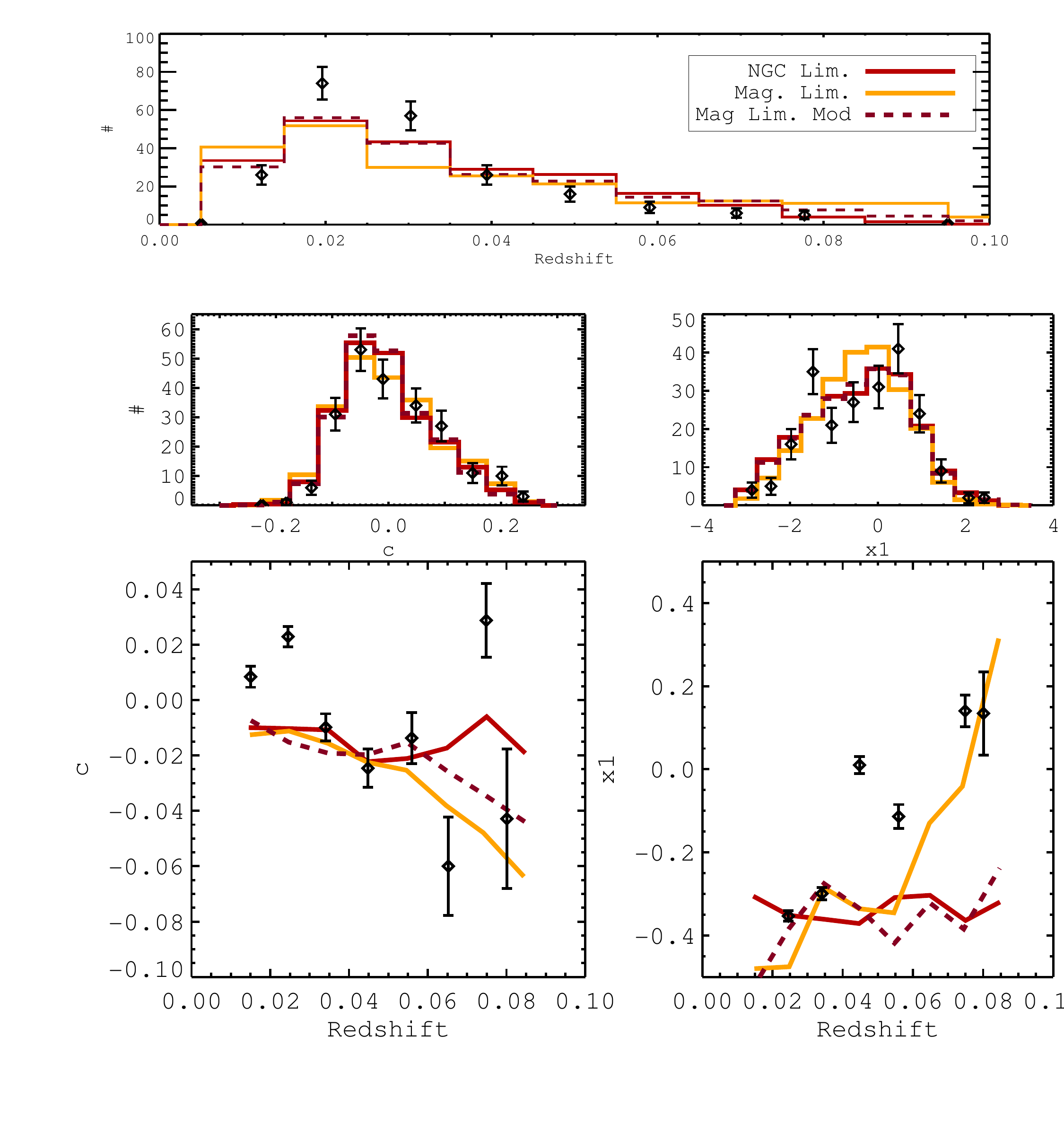}
\caption{Comparison of distributions for low-z data (points) and MC (histogram), where each MC distribution is scaled to have the same sample size as the data.  Similar to Plot~\ref{fig:simverify}.  Here we simulate three scenarios: a NGC-limited survey, a magnitude-limited survey, and a modified magnitude-limited survey to show the error on the cumulative selection bias.}
\label{fig:ngc}
\end{figure}

For the low-z sample, it is much more difficult to simulate the individual subsamples because of the smaller statistics, discoveries that are external to the survey, and in some cases, multiple telescopes used for observations.  Selection effects in these surveys result from the discovery threshold and the selection for spectroscopic follow-up, not from limitations in SNR, because the SNR reach $\sim10$ even for the faintest observations in the sample.  To quantify the Malmquist bias in the low-z sample, we find the spectroscopic/selection efficiency of the aggregate of the low-z surveys.  Similar to how we found the spectroscopic efficiency of the PS1 survey, we find the selection efficiency function that best describes the combined low-z sample: 
\begin{equation}
\textrm{Spec. Sel. Eff.} = e^{-\frac{(m_B-14)}{1.5}}
\end{equation}
where $m_B$ is the peak B band magnitude and is greater than 14.  This efficiency function assumes a flat volumetric rate at low-z.  A comparison of the simulation to the data, including trends of the SALT2 light curve parameters color ($c$) and light-curve shape ($x_1$) with redshift, is shown in Fig.~\ref{fig:ngc}.  For the entire low-z sample, we find a mean difference in colors for $z>0.04$ and $z<0.04$ of $\Delta c = -0.031\pm0.011$ and a mean difference in light-curve shape of $\Delta x_1=+0.53\pm0.16$.  We analyze these same trends for each low-z subsample in Appendix B.  The color and stretch trends with redshift appear to change more significantly in CfA3 and CfA4 than in JRK07 and CSP.

We further probe whether the survey is magnitude-limited in Appendix B by comparing the redshift distribution of the low-z sample to the distribution of galaxy redshifts in the New General Catalogue (NGC).  We find that the redshift distribution of the SNe is actually fairly well represented by the redshifts in the NGC, though higher redshift galaxies are slightly over-represented.  This suggests that any preference for brighter targets at a given redshift should be negligible.  Rather, the only selection bias of the low-z sample should be the selection bias of the NGC galaxies themselves.
\
In Fig.~\ref{fig:ngc}, we compare simulations of a NGC-based survey, where the redshift distribution matches the NGC distribution, to the data.  The redshift distribution of the simulations of the magnitude limited and galaxy-targeted surveys are similar, though the NGC-based simulation cannot replicate the trends of color and stretch with redshift.  Therefore, like we did for the PS1 selection bias, we determine the systematic uncertainty on the selection bias empirically rather than using the NGC-follow-up to determine the systematic uncertainty.  A survey that reproduces the trends with color and stretch less well than our optimal method by $2\sigma$ is also shown in Fig.~\ref{fig:ngc}.

This uncertainty for the low-z sample is roughly $0.003$ mag for the entire low-z sample.   The uncertainty due to generalizing all of the low-z surveys is included in the overall selection effect uncertainty in Table~\ref{tab:dom}.  The change in $w$ from not accounting for any low-z Malmquist bias is large: $\Delta w \approx - 0.035$ (SN-only).  By opting to determine the error on the selection bias empirically rather than allowing for a systematic based on a model in which there is no selection bias, we significantly reduce the systematic uncertainty of the Malmquist bias.

\section{Light Curve Modeling}
\label{sec:lcfits}

To optimize the use of SN\,Ia as standard candles to determine distances, the majority of the SN\,Ia light curve fitters (e.g MLCS2k2; \citealp*{2007ApJ...659..122J}, SALT2; \citealp{2007A&A...466...11G}, SiFTo; \citealp{2008ApJ...681..482C}, CMAGIC; \citealp{2009ApJ...699L.139W}, SNooPY; \citealp{SNooPy}, BAYESN; \cite{Mandel/etal:2011}) include two corrections to the observed peak magnitude of the SN: one using the width/slope of the light curve and the other using the color of the light curves.  Here we use SALT2 to fit the light curves, and determine linear relations between luminosity with light curve width and color according to the Tripp distance estimator.  We analyze the systematic uncertainties in the following section.  

\subsection{SN\,Ia Color}

SN\,Ia light curve fit parameters determined using the SNANA's light curve fitter with a SALT2 model \citep{2010guylightcurves} are presented in R14.  Recently,  S13 found that a model of the intrinsic brightness variation of SN\,Ia that is dominated by achromatic variation may lead to a similar color-luminosity relation as one in which there is a large amount of chromatic variation.  Here, we find the different values of $w$ when the dispersion of SN\,Ia distances is attributed to a model that contains mostly ($\ge 70\%$) luminosity variation or a model that contains mostly color variation.

R14 shows the values of the color-luminosity relation, $\beta$, depend on different assumptions about the source of intrinsic scatter.  They find that $\beta=3.10\pm0.12$ when scatter is attributed to luminosity variation and $\beta=3.86\pm0.15$ when scatter is attributed to color variation.  The latter value is only $<2\sigma$ from a MW-like reddening law, though R14 notes this high value is strongly pulled by the low-z sample only.  To understand the consequences of these two assumptions about intrinsic scatter, we first match simulations, as explained in the previous section, of two variation models to the data.  We create simulations using SNANA with two models, one in which the majority of scatter is due to luminosity variation (called `Guy10', \citealp{2010guylightcurves}, color/luminosity variation - $30\%/70\%$) and the other in which the majority of scatter is due to color variation (called `Ch11'-\citealp{Chotard_thesis}, color/luminosity variation - $75\%/25\%$).  For the model with a majority of luminosity variation,  the color-luminosity relation, $\beta$ is set to be $3.1$, the value found after attributing all of the Hubble residual scatter to luminosity variation (R14).  For the model with mostly color variation, SN\,Ia color is composed of a dust component that correlates with luminosity via a Milky-Way-like reddening law  ($\beta=4.1$) and a variation component \citep{Chotard_thesis}.  SNANA provides these two models.  For \cite{Chotard_thesis},  SNANA converts a covariance matrix among bands into a model of SED variations.  The \cite{Chotard_thesis} model used is denoted `$\textrm{C11}_0$' in SNANA.  

\begin{figure}[h]
\centering
\epsscale{\xxScale}  
\plotone{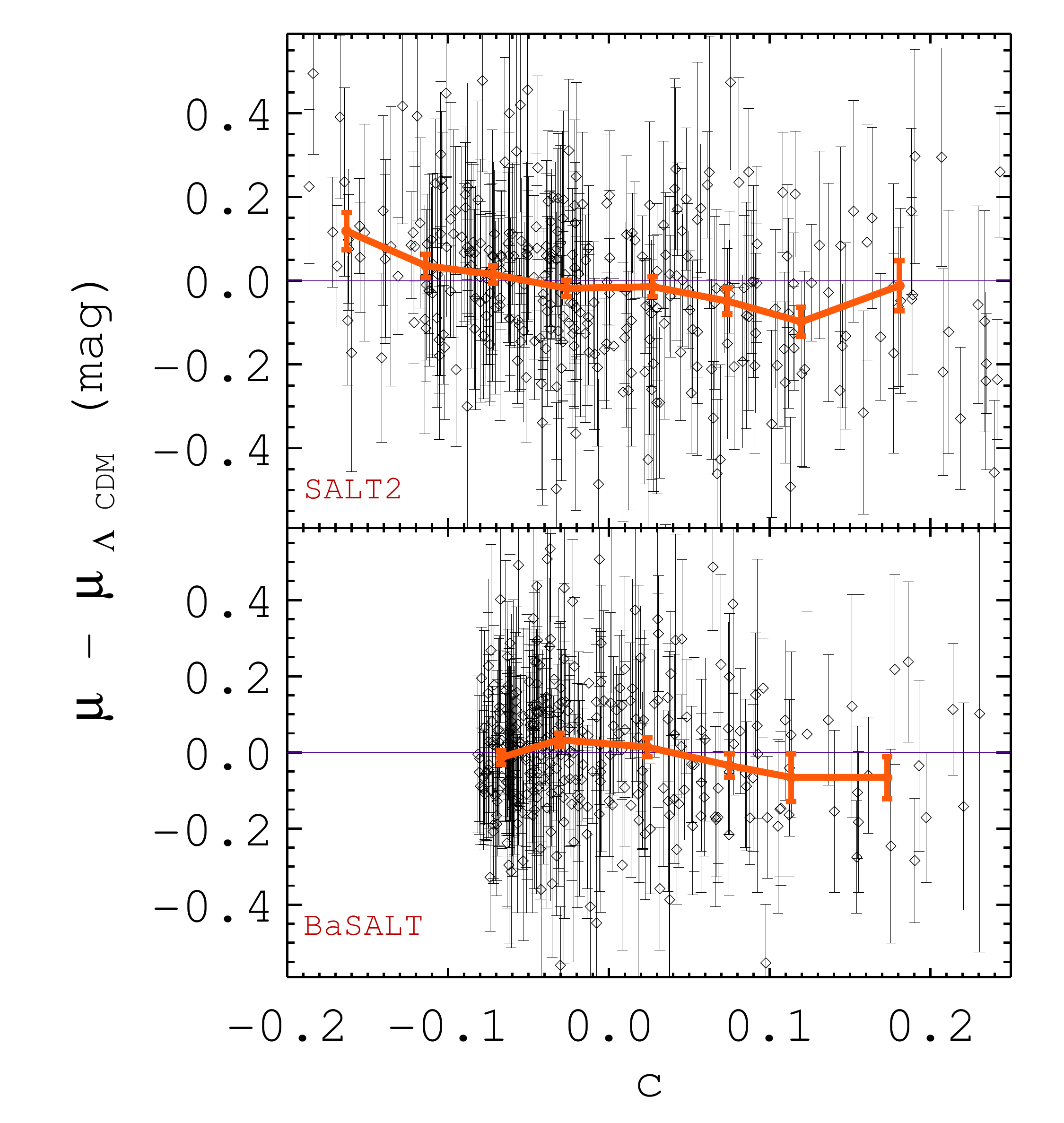}
\caption{Hubble Residuals versus color for the SALT2 (top) and BaSALT (bottom) methods over the entire redshift range.}
\label{fig:baltps1}
\end{figure}

S13 showed that trends in Hubble residuals versus color depend not only on the intrinsic scatter but also the underlying color distribution.  Parameters for the underlying color and stretch distributions that best match simulations to the data are given in Table~\ref{tab:x1c}.  These values are found using a grid-based search of the $x_1$ and $c$ asymmetric gaussian parameters.  S13 showed that simulations with a Guy10 variation model, combined with a slightly asymmetric underlying color distribution \citep{kessler_13_var}, cannot reproduce the significant asymmetry around $c\sim-0.1$ seen in the trends of Hubble residuals versus color (similarly shown for PS1+lz sample, given in Fig.~\ref{fig:baltps1} - top).  Some correlation between color and Hubble residuals should be expected from the Tripp distance estimator, though the trend seen here can be best explained by a narrow and asymmetric underlying color distribution (Table~\ref{tab:x1c}). To improve the consistency of the Guy10 model with observations, we find that the underlying color distributions for this variation model should be significantly asymmetric.  The distribution presented here is much more asymmetric than that given in \cite{kessler_13_var} for the SDSS or SNLS surveys.  
 
\begin{figure}[h!]
\centering
\epsscale{\xxScale}  
\plotone{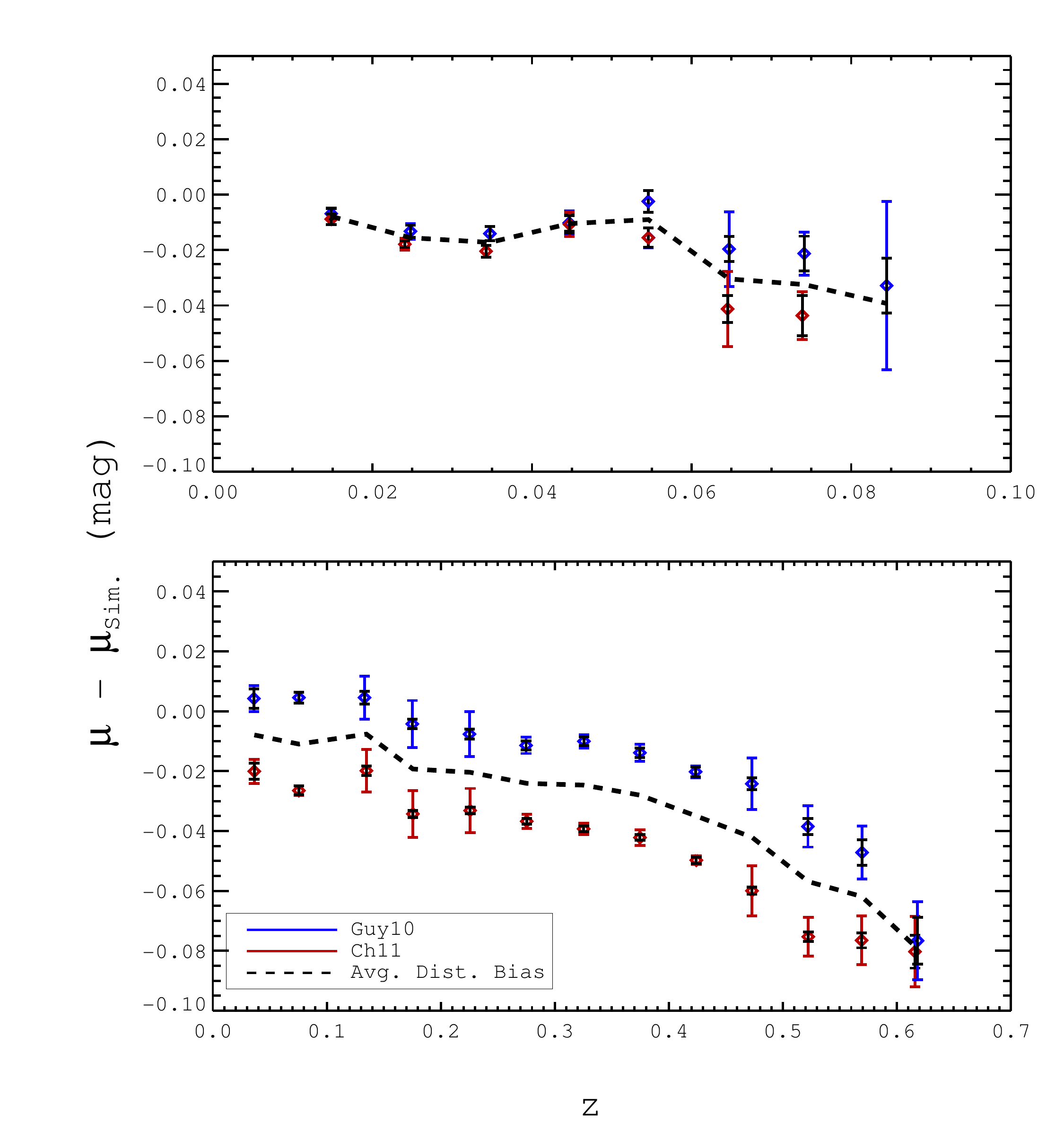}
\caption{The distance biases over the entire redshift range for our two variation models.  The average of these biases is also shown.  The inner errors show the errors from the simulation, whereas the outer errors include this error as well as the error in the selection bias uncertainty.}
\label{fig:malm3}
\end{figure}

\begin{figure}[h!]
\centering
\epsscale{\xxScale}  
\plotone{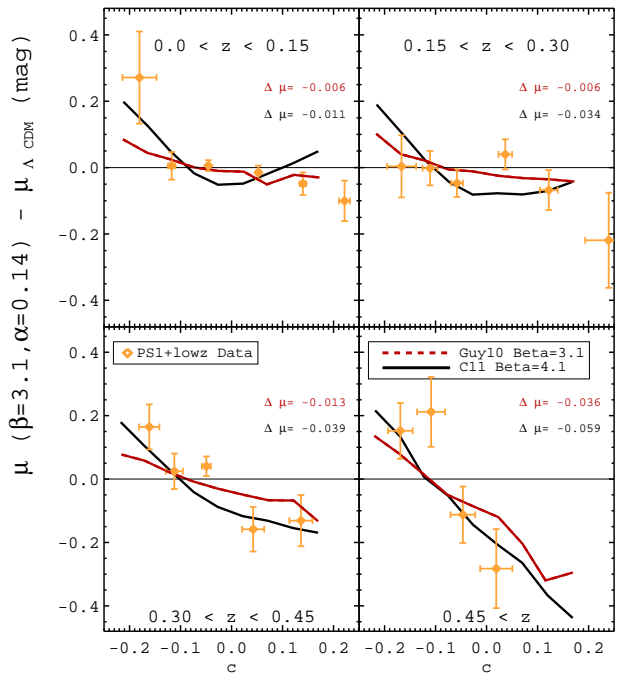}
\caption{Hubble Residuals as a function of color for different redshift bins.  Distances are found using the conventional SALT2 method where residual scatter is attributed to luminosity variation.  One simulation is in accordance with the conventional SALT2 assumptions, whereas the other simulation follows the color variation model (BaSALT).  The lowest redshift bin contains all of the Nearby supernovae. }
\label{fig:bcolor}
\end{figure}

We present our distances biases with redshift in Fig.~\ref{fig:malm3}.  Distances included here are found once intrinsic dispersion of SN\,Ia is attributed to luminosity variation.  We find differences in the distance corrections to be up to $0.03$ mag and note that the offsets are fairly constant with redshift for the PS1 sample.  We find overall a mean offset of $-0.01$ mag for both models with the low-z sample, and this offset is subtracted out from both the low-z and high-z samples as they are combined.  This offset is due to the asymmetric underlying color distribution and covariances between $m_b$ and $c$.  We further understand the predictions by our two models by analyzing trends between Hubble residual and color at separate redshift bins in Fig.~\ref{fig:bcolor}.  While there are discrepancies in the predictions for the two models of SN variation, the statistics of the PS1+lz sample do not favor either model; the data cannot break the degeneracy.   In Fig.~\ref{fig:simverify}, we also overplot a simulation with our color variation model, and find no noticeable differences from our simulation with luminosity variation.  

Recent analyses (e.g., \citealp{campbell_13_sdss}, \citealp{kessler_13_var}) attempt to remove any fitter bias (including the Malmquist bias) by using simulations to find the mean distance residual for a given redshift, like the ones we have done here.  We may use our two different simulations to find the systematic uncertainty in our distance corrections from our incomplete understanding of the true variation model.  For our primary analysis, we correct for the average distance residual at all redshifts from the two simulations.  These average distance residuals are shown in Fig.~\ref{fig:malm3}. The difference in $w$ after applying the corrections from one model or the other is $\Delta w\approx0.055$ (Lum.-Col.).  By taking the average, we reduce the systematic uncertainty due to the color models by a factor of 2.

\begin{deluxetable}{lll | ccc  }[h!]
\tablecaption{Asymmetric Gaussian parameters to describe the parent distribution of $x_1$ and $c$.} 
\tablehead{
\colhead{Parameter} &
\colhead{Sample} &
\colhead{Intr. Variation} &
\colhead{$\bar{x}$} &
\colhead{$\sigMINUS$ } &
\colhead{$\sigPLUS$ }  \\
}
\startdata
  
    $c$ &  PS1 & Ch11   &  $-0.1$    &  0.0  & 0.095  \\
        $c$  & PS1 & Guy10   &  $-0.08$    &  0.04  & 0.13  \\
   $c$  & Low-z & Ch11  &  -0.09    &  0.0  & 0.12  \\
   $c$  & Low-z & Guy10   &  -0.05    &  0.04  & 0.13   \\
   
   $x_1$ & PS1 & Ch11 &  0.5   &  1.0   & 0.5       \\
      $x_1$ & PS1& Guy10  &  -0.3   &  1.2   & 0.8        \\
      $x_1$ & Low-z & Ch11  &  0.5    &  1.0   & 0.5        \\
   $x_1$ & Low-z & Guy10  &  -0.3    &  1.2   & 0.8        \\

\enddata
\tablecomments{The parameters defining the asymmetric Gaussian for the color and light-curve shape distributions: $e^{ [-(x - \bar{x})^2/2\sigMINUS^2] } $ for $x < \bar{x}$ and $e^{ [-(x - \bar{x})^2/2\sigPLUS^2] }$ for $x > \bar{x}$.  The optimized parameters for the variation models with a majority due to color variation (Ch11-Chotard et al., color/luminosity variation - 75\%/25\%) and luminosity variation (Guy10-Guy et al., color/luminosity variation - 30\%/70\%) are given.}
  \label{tab:x1c}
\end{deluxetable} 

A separate way to determine the systematic uncertainty from SN color is to compare the values of $w$ when using SALT2 in the conventional manner (attribute intrinsic scatter to luminosity variation) as well as using BaSALT (S13), a Bayesian approach that separates the color of each SN into components of color variation and dust. To retrieve the component of color that correlates with luminosity ($c_{\textrm{dust}}$), BaSALT applies a Bayesian prior to the observed color ($c_{\textrm{obs}}$) such that
\begin{equation}
c_{\textrm{dust}}=\frac{1}{P} \int_{c>\bar{c}} c e^{-(c-c_{\textrm{obs}})/2\sigma_{c_n}^2}e^{-(c-\bar{c})^2/\tau_S(z)^2} \partial c.
\label{eqn:BALT}
\end{equation}

Here, $c_n$ is the noise from the color measurement and $P$ is a normalization constant.  The second part of Eqn.~\ref{eqn:BALT} describes the Bayesian prior \citep{riess96} for the dust distribution where $\bar{c}$ is the blue cutoff of the distribution.  ${\tau_S(z)}$ describes the shape of the one sided Gaussian due to extinction for a given redshift $z$ for each survey $S$; the dependence of $\tau$ on survey and redshift allows selection effects to be modeled.  As discussed in the previous section, for the low-z sample we do not expect selection effects to significantly vary with redshift and therefore we do not change $\tau$ with redshift.  For the PS1 sample, however, $\tau$ varies with redshift\footnote{For Low-z: $\tau=0.11$.  For PS1: $\tau={[0.11,0.10,0.08,0.06]}$ for $\vec{z}={[0.1,0.3,0.5,0.7]}.$}.  
The results are shown in Fig.~\ref{fig:baltps1} (bottom).  The BaSALT method does not allow the color fits to be bluer than $c=-0.1$ and therefore the particular non-linearity between Hubble residual and color is much weaker with this method.  However, given the BaSALT method, we introduce with the color prior additional correlations between color and distance that are dependent on the uncertainty of the prior.  Therefore, we only use the BaSALT method as a way to confirm our results using the simulations discussed above.  We find a change in $w$ of $\Delta w=w_B-w_S=-0.06$ when the BaSALT method is used instead of the SALT2 method.  This is nearly equal to the difference found for $w$ from using distance corrections from simulations of the different scatter models.  With the BaSALT method, we still must use simulations to correct for any further biases due to covariances between color and the other light curve fit parameters.  If information about the underlying color distribution was included during the light curve fit itself, this would not be needed.

The difference in recovered cosmological parameters due to variation models largely depends on the mean color of the sample.  If the mean color of the sample is constant with redshift, we should not expect significant ($|\Delta w| > 0.01$) discrepancies due to finding the wrong $\beta$.  In the PS1+lz sample, the PS1 subsample is bluer by $\sim0.03$ mag than the low-z sample, and thus the effect of an incorrect $\beta$ is significant.

\subsection{Non-linear Light-curve Shape}

We now explore whether the relation between the light-curve shape and luminosity is adequately described by a linear model.  Similar to the analysis of color, we observe the trend between Hubble residuals and light-curve shape, known in SALT2 as `stretch' ($x_1$).  Doing so, we find that a second-order polynomial ($\alpha_1 \times x_1 + \alpha_2 \times x_1^2$, where $[a_1,a_2]=[0.160\pm0.010, 0.017\pm0.007]$) appears to better fit the data than the conventional linear model ($\alpha \times x_1$).  The reduction in $\chi^2$ when including the second order is from $312$ to $301.8$ for 312 SNe (after uncertainties have been inflated such that $\chi^2_\nu=1$ and not including the $\alpha_2$ term in the uncertainty).    The significance found here is larger than that found in Sullivan et al. 2010 or Sullivan et al. 2011 when they examine a second order stretch correction.  In the Sullivan et al. papers, they find two $\alpha$ values for high and low-stretch values, both of which were larger than the value of $\alpha$ found for the whole sample.  We find a discrepancy in Hubble residuals for high ($x_1>0.5$) and low stretch ($x_1<0.5$) values to be $\Delta \mu = 0.042 \pm 0.020$ (after Malmquist correction).

\begin{figure}[h]
\centering
\epsscale{\xxScale}  
\plotone{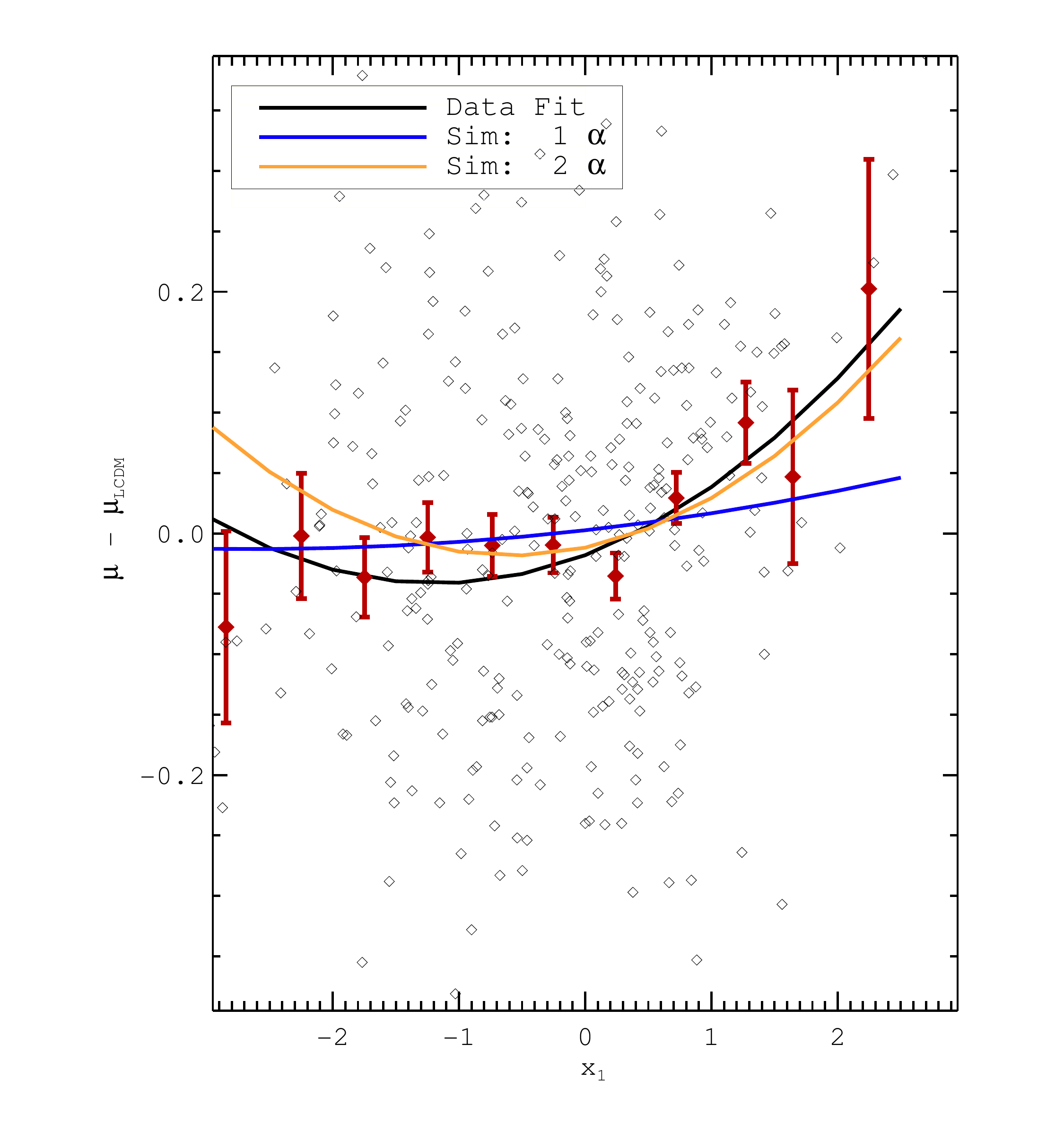}
\caption{Hubble Residuals versus stretch for the PS1+lz data and simulations.  The results for simulations are from a 1 $\alpha$ and 2 $\alpha$ model.}
\label{fig:phillips}
\end{figure}

A more practical approach to understand the source of the second-order trend of Hubble residuals with stretch is to reproduce the observed effect in simulations.  We find (Fig.~\ref{fig:phillips}) that the trend seen in the data may be replicated with two different $\alpha$ parameters for high ($x_1 > 0$) and low ($x_1<0$) stretch values.  We use this split function as we do not yet have the tools to simulate a second order polynomial of the stretch-luminosity relation.  A comparison of results of various simulations with the data is presented in Fig.~\ref{fig:phillips}.  We show that the quadratic trend between Hubble residuals and stretch that is seen in the data is not seen in simulations with only one $\alpha$ ($\alpha=0.14$).  However, for the $2\alpha$ model, where for $x_1<0$, $\alpha=0.08$ and for $x_1>0$, $\alpha=0.17$, the quadratic trend is observed.  More work must be done to determine if this effect is real, and if a continuous, but non-linear, Phillips relation is empirically favorable to the discontinuous relation presented here.  We find the change in the determined value of $w$ when accounting for a second-order light-curve shape correction is small: $\Delta w = +0.01$.  Since there is only mild evidence ($\sim 2\sigma$) that a second-order light curve shape correction is beneficial, it is not included in our overall uncertainty budget.

\subsection{Host Galaxy Dependence}
\nobreak
\label{sec:hosgalaxy}

We examine here whether Hubble residuals of the SNe in the PS1+low-z sample correlate with properties of the host galaxies of SNe.  These correlations have been shown for other SN\,Ia samples in multiple recent studies (e.g., \citealp{2010ApJ...715..743K}, \citealp{2010MNRAS.406..782S}, \citealp{2010ApJ...722..566L}).  While age, metallicity, and Star-Formation-Rate (SFR) of host galaxies all also have been observed to correlate with Hubble Residuals (e.g., \citealp{gupta2011}, \citealp{2013ApJ...764..191H}), here we focus on the masses of the host galaxies.  

To determine the masses of host galaxies of PS1 SN\,Ia, we combine PS1 observations with $u$ band data from SDSS.  Since the PS1 Medium Deep Survey has now observed each of the SN\,Ia host galaxies for over two years, we stacked deep (SNR of 10 at $\rps \sim 24$) SN-free templates and used SExtractor's  FLUX\_AUTO \citep{Sex} to determine the flux values in \gps,\rps,\ips,\zps,\yps.  The measured magnitudes are analyzed with the SED fitting software Multi-wavelength Analysis of Galaxy Physical Properties \citep{daCunha2011} to calculate the stellar masses of host galaxies.  To verify both our galaxy photometry and the mass fitting routine, we ran our mass fitter on PS1 photometry of host galaxies in the \cite{gupta2011} sample, and found differences with \cite{gupta2011} to be $<log (M_{\odot})\sim 0.05$.

\begin{figure}[h!]
\centering
\epsscale{\xxScale}  
\plotone{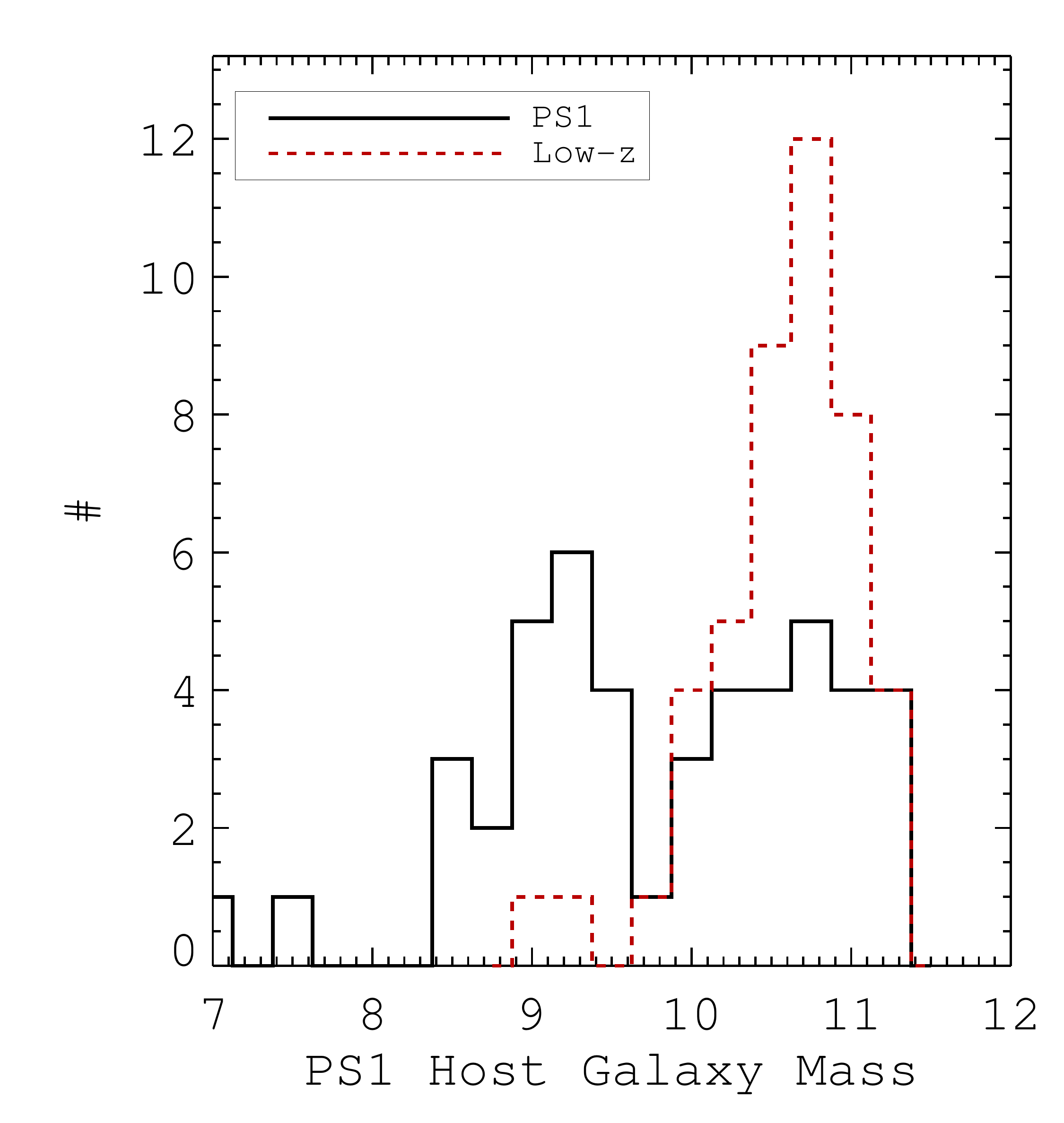}
\caption{The host galaxy mass distribution for the PS1 sample and combined low-z sample. }
\label{fig:hosthist}
\end{figure}

\begin{figure}[h!]
\centering
\epsscale{\xxScale}  
\plotone{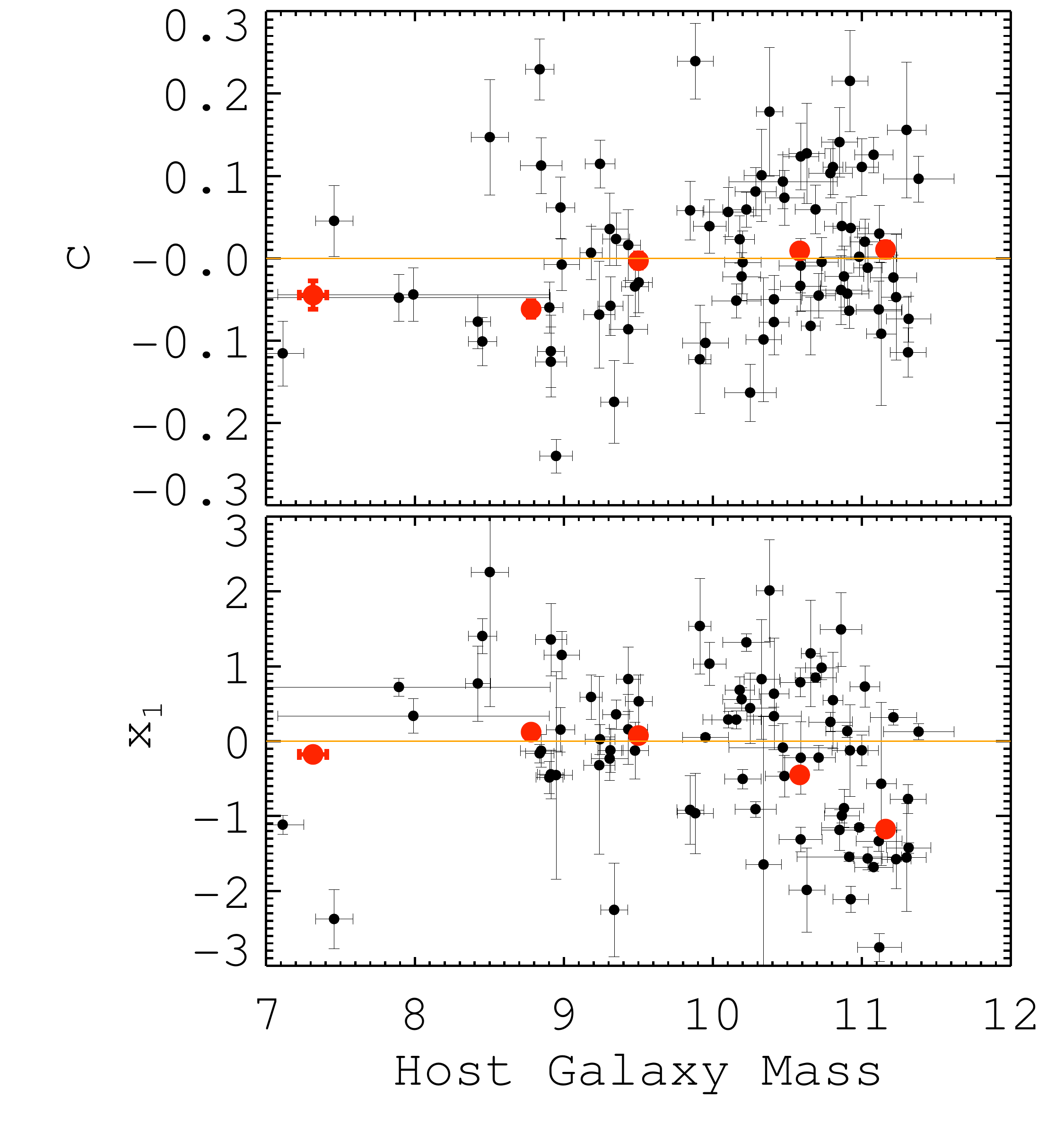}
\caption{The SALT2 stretch parameter $x_1$ and color $c$ as a function of the host galaxy $M_{stellar}$ for the entire PS1 sample.  The red points show the weighted means.  }
\label{fig:hoststretch}
\end{figure}

The mass distributions of the host galaxies from the combined PS1+low-z sample analyzed for this paper are shown in Fig.~\ref{fig:hosthist}.  We currently have masses for 110 of the \SNIAPSused~PS1 host galaxies and 61 of the \SNIAlowzused~low-z host galaxies.  The host galaxy masses of the low-z sample are from \cite{2010MNRAS.406..782S} and the incompleteness limits a full analysis.  Because most low-z supernovae were found in catalogued galaxies, and catalogued galaxies are generally brighter and more massive on average, the host galaxies in the low-z surveys are $log(M_{\odot})\sim0.5$ more massive than those in the PS1 survey.  We find from Fig.~\ref{fig:hoststretch} a trend such that supernovae with more massive host galaxies have lower stretch values, though quantifying this trend is obscured by a number of supernovae with very low stretch errors ($dx_1<0.1$).  We find a relation between Hubble residual and color with a slope of $0.0054\pm0.0049$.  While the significance of this relation is weak, it is in agreement with the trend observed by \cite{Ch13}.

\begin{figure}[h!]
\centering
\epsscale{\xxScale}  
\plotone{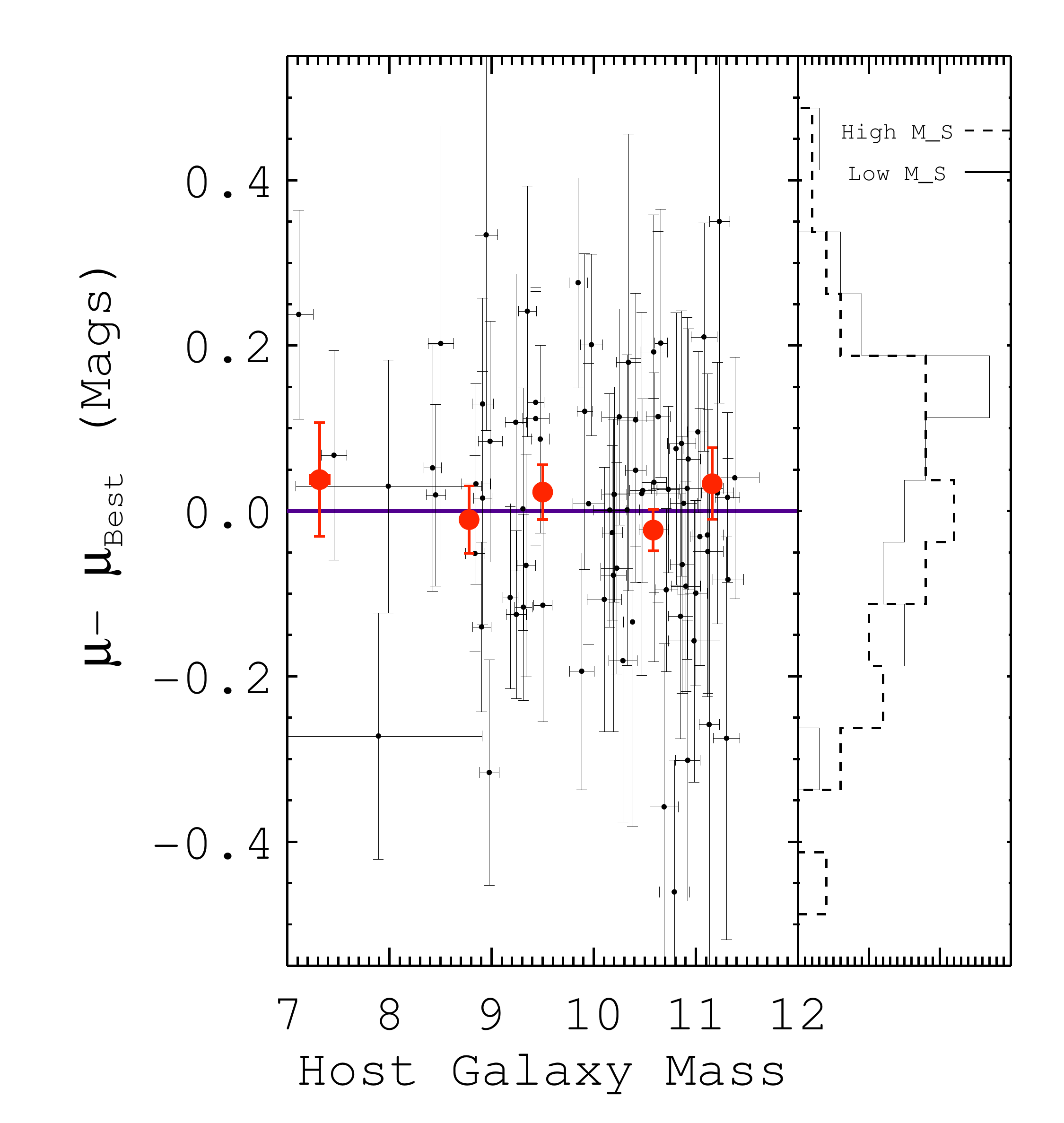}
\caption{Residuals from the best-fitting cosmology for PS1 as a function of host galaxy $M_{stellar}$.  The red points show the weighted mean residuals in bins of host $M_{stellar}$ }
\label{fig:hosthub}
\end{figure}

In Fig.~\ref{fig:hosthub}, we show the trend between Hubble residuals and the host galaxy mass.   We find a step-size in Hubble residuals, for a mass split of $\textrm{log}(M_{\odot})=10$, of $\Delta \mu = 0.037\pm0.032$ mags.  While this is consistent with the step-size seen in previous studies  ($\sim0.08$ mag; e.g., \citealp{2010MNRAS.406..782S}, \citealp{gupta2011}), it is also consistent with zero.  For the PS1 SNe alone, we find only a difference of $0.019\pm0.025$, weaker than the typical value seen.  Part of the impact of adding in the low-z sample may be due to the fact that low-z hosts have higher masses, on average, than high-z hosts.  Therefore, any difference in luminosity is degenerate with discrepancies in calibration for the low-z and high-z samples.  It is also possible that the low value of intrinsic scatter seen in the PS1 sample ($\sigma_{int}=0.07$) found for the PS1 sample is related to the low significance of the host galaxy relation.  One concern is that the mass fits are likely subject to selection biases as the number of observations in separate filters decreases at higher redshift, (e.g., Galex and UKIDSS detect more nearby hosts).  However, \cite{Smith2012} finds discrepancies in the determination of mass for SNe with high and low redshift to be negligible for the SDSS+SNLS sample.  

As we do not understand why the difference in luminosity between low and high-mass galaxies is weaker in the PS1+lz sample than in other samples, we include as a systematic uncertainty the possibility that the host galaxy relation is as seen in other surveys (e.g. \citealp{Ch13}) versus the size seen in our sample.  We implement for the PS1 sample the same mass-distribution split done in  \cite{2010MNRAS.406..782S} at ($\textrm{log}(M_{\odot})=10$) and apply a difference in luminosities between SNe with high and low massive host galaxies at $\sim0.075$ mag \citep{2010MNRAS.406..782S}.  When we do not have sufficient host galaxy photometry to find masses, we do not correct the SN distance.  If, however, the host is not visible, we assume the mass is low.  We find that the correction based on host galaxy masses changes our value of $w$ by $\Delta w = 0.026$ (SN+BAO+CMB+H0: host correction - no-host correction).  The overall uncertainty for the host galaxy correction is shown in Table~\ref{tab:dom}.  Its effect on the recovery of cosmological parameters is not as large as some of the other uncertainties, which may be partly due to the incompleteness of host galaxy measurements for the low-z sample.

\section{Redshift Uncertainties}
\nobreak
\label{sec:pecvel}

Inhomogeneities in the density field generate peculiar velocities that perturb the Hubble flow (see \citealp{StraussWillick1995} for a review).  The overall uncertainty in redshift for each SN can be expressed as:
\begin{equation}
\sigma_z^2 = \sigzspec^2 + \sigzflow^2.
\end{equation}
where $\sigzspec$ is the spectroscopic uncertainty and $\sigzflow$ is the correlated velocity uncertainty of SN peculiar velocities relative to the Hubble flow. 

These uncertainties propagate towards an uncertainty in distance by \citep{2009ApJS..185...32K}:
\begin{equation}
  \sigmudz =  \sigma_z \left(\frac{5}{\ln 10}\right) \frac{1+z}{z(1+z/2)} ~.
  \label{eq:sigmudz}
\end{equation}
The first source of redshift uncertainty, $\sigzspec$, is either from the supernova spectrum ($\sigzspec=0.005$, Foley et al. 2009) or the host-galaxy spectrum ($\sigzspec=0.0005-0.001$).  We update all nearby redshifts from the NASA/IPAC Extragalactic Database\footnote{http://ned.ipac.caltech.edu/}.  Before analyzing local flow corrections, we transform all supernova redshifts into the comoving frame of the CMB  \citep{Fixsen96}.  

Large-scale flows are correlated across the nearby Universe and the peculiar velocities are not another form of random noise.  To correct for the correlated flows of local SNe ($z<0.10$), we follow \citet{2007neillpecvel}.  Their model is based on the galaxy density field in the nearby universe ($z <0.06$), and includes infall into nearby superclusters and large-scale bulk flow \cite{2004MNRAS.352...61H}. The large-scale residual bulk flow arising from the density field at larger distances is at $150 \pm 43~\textrm{km/s}$ in the direction $l = 345^{\circ} \pm 20^{\circ}$, $b = 8^{\circ} \pm 13^{\circ}$ \citep{2012MNRAS.420..447T}. The accuracy of the corrections is estimated to be $ \sigzflow =150$ km $\mathrm{s}^{-1}$.   We do not correct for further covariances between supernovae velocities, though large sample of low-z supernovae spread out across the sky should mitigate the effects of any further correlated flows that are not included in the model \citep{davis2011}.  We find that not including the coherent flow corrections significantly increases the $\chi^2$ of the distance residuals from $\chi^2=313.0$ to $\chi^2=332.0$.

\begin{figure}[h!]
\centering
\epsscale{\xxScale}  
\plotone{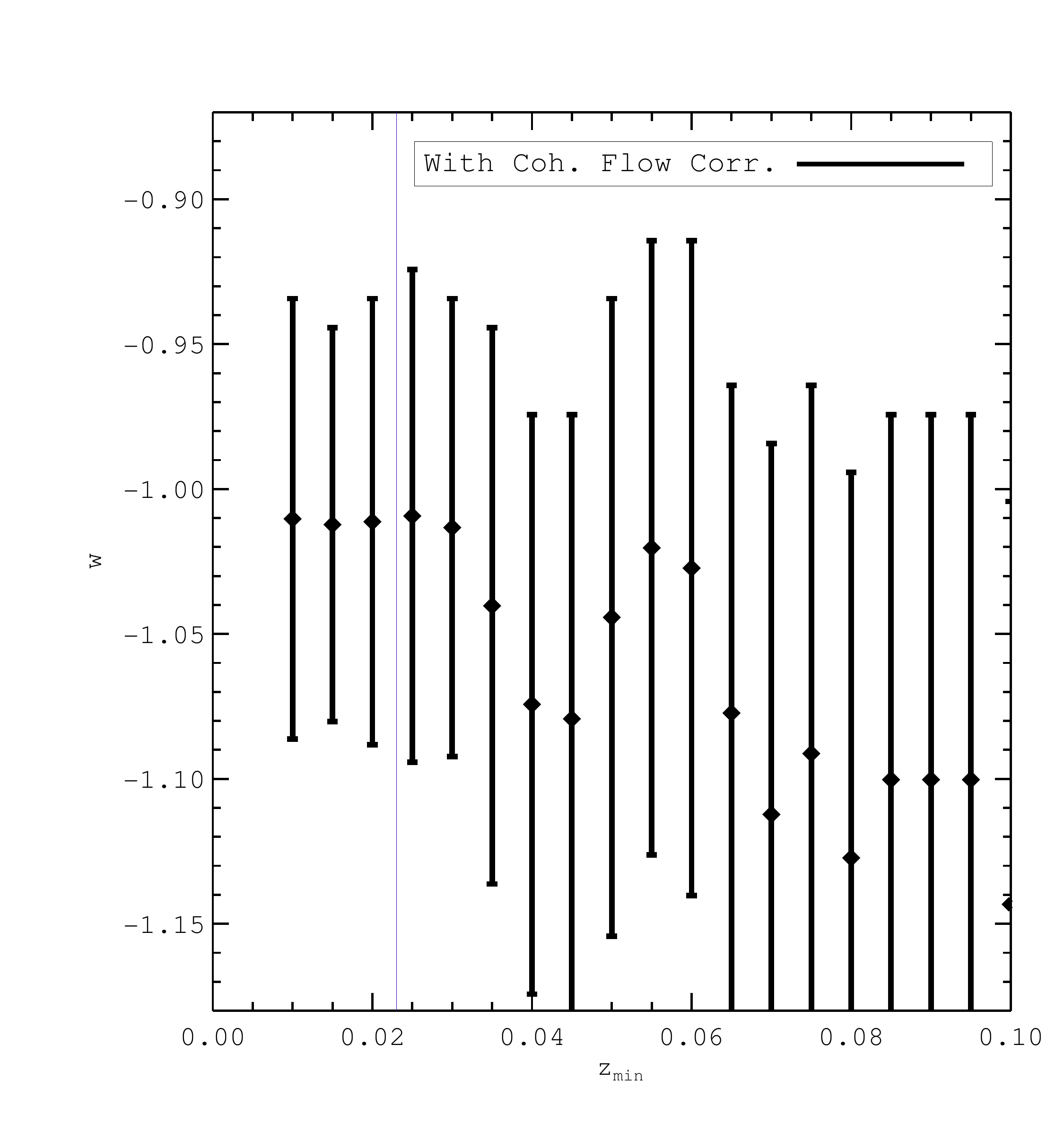}
\caption{Variation in dark energy equation of state parameter $w$ when the minimum redshift $z_{\textrm{min}}$ of the sample is varied.  A vertical line at $z=0.023$ is shown.}
\label{fig:peculiar}
\end{figure}

To confirm that the coherent flow corrections remove a a potential bias due to our possible location in an underdense region of the Universe (\citealp{zehavi98} and \citealp{2007ApJ...659..122J}), we analyze how $w$ varies with the minimum redshift of the sample. Past evidence (e.g., \citealp{2007ApJ...659..122J}) showed that the Hubble constant is larger within $z\sim0.023$, though this claim has been shown to depend on assumptions about the SN reddening law \citep{conley07}.  For the combined low-z+PS1 sample, the variation of $w$ with minimum redshift is shown in Fig.~\ref{fig:peculiar}.  Here we see a change of $ \Delta w  = +0.012$ after coherent flow corrections.  This change appears to be consistent with the claim in C11 that any previously seen change in $w$ when varying the minimum redshift may be due to shot noise.  To be conservative, we still include this effect as a systematic uncertainty.  The covariance matrix presented in \S\ref{sec:lcfits} does not properly allow for the removal of SNe from the original sample (from changing minimum redshift).  Therefore, we find the difference in the means of the distance residuals for the two samples ($0.01<z<0.10,0.023<z<0.10$) to be $-0.004$ mag and add this as our $\Delta \mu$ in Eq.~\ref{eqn:covar}.  The effects of this uncertainty is included in Table~\ref{tab:dom} and is one of the smallest systematic uncertainties.  We note in Fig.~\ref{fig:peculiar} that the steep change in $w$ around $0.035$ is also likely due to shot noise.  For this minimum redshift, the total low-z sample is reduced from $\sim190$ SNe to $\sim60$.

\section{Milky-Way Extinction Correction}
\label{sec:extinction}

For each SN, we use an estimate of the dust along the line of sight calibrated by  \cite{1998ApJ...500..525S} (hereafter, SFD98).  The correction applied for MW extinction is more important for the low-z sample as the PS1 Medium Deep fields were chosen to have low extinction whereas the low-z supernova were discovered all over the sky.  In most SN papers (e.g., \citealp{2009ApJS..185...32K}), the extinction law is given by \cite{ccm89}.  Recently, \cite{Schlafly11} analyzed the colors of stars with spectra in the Sloan Digital Sky Survey and found that their results preferred a \cite{Fitz99} reddening law with $R_V=3.1$ to that of \cite{ccm89}.  They also find that SFD98 systematically overestimates $E(B-V)$, and that moreover the true reddening $E(B-V)$ is somewhat nonlinear in $E(B-V)_{SFD}$ such that\footnote{see Fig. 8 of  \cite{Schlafly11}.}
\begin{multline}
E(B-V)=0.94 \times E(B-V)_{SFD}  \textrm{ : }E(B-V)_{SFD} <0.1 ,\\
E(B-V)=0.86 \times E(B-V)_{SFD}  \textrm{ : } E(B-V)_{SFD} >0.1 .
 \label{eqn:MW}
\end{multline}
Here, we implement both of these changes.  With these corrections, the value of $w$ found (SNe only) is changed by $\Delta w = + 0.026$.  The effect of this correction depends on the mean extinction values of the low and high-z subsamples.   The mean MW E(B-V) value for the PS1 sample is $\sim0.02$ while for the Nearby samples it is $\sim0.07$.

The uncertainty in SFD, besides the random uncertainty in the $E(B-V)$ values, is difficult to quantify.  In the construction of the SFD map, the largest uncertainty is in the subtraction of the zodiacal light, which has a complex spatial dependence on the sky.  \cite{peek10} finds errors in SFD of $0.005$ mag rms over the high Galactic latitude sky, averaged over an angular scale of 3 degrees.  Following this work, we adopt a conservative additive uncertainty of $0.01$ mag stemming from the problematic zodiacal light subtraction.  There are also multiplicative systematic uncertainties in the temperature correction and the conversion from 100 micron depth to $E(B-V)$. While \cite{Schlafly11} have determined the average conversion to an accuracy $<1\%$,  there are still $\sim10\%$ coherent errors in few-degree patches over the sky because of the variance of dust properties and uncertainties in the temperature correction.  Also, as seen in Eqn.~\ref{eqn:MW}, there are non-linearities to the SFD correction.  In total, we accept a $10\%$ scale uncertainty to the corrected SFD values.  Finally, while the temperature correction most likely causes the best fit normalization of the dust map to vary over the sky, the dust extinction spectrum is relatively stable, and at least in regions with $ E (B - V ) <  0.5$, objects can be dereddened in the optical assuming a universal extinction law to within a few percent accuracy.  Therefore, we adopt as our systematic uncertainty $0.015 + 0.10 \times E(B-V)$ and cut off all SNe with $E(B-V)$ values greater than 0.5.  The overall effects of the extinction uncertainties are included in Table~\ref{tab:dom}.

\section{Overall Review and Discussion}
\label{sec:system}
\begin{figure}[h!]
\centering
\epsscale{\xxScale}  
\plotone{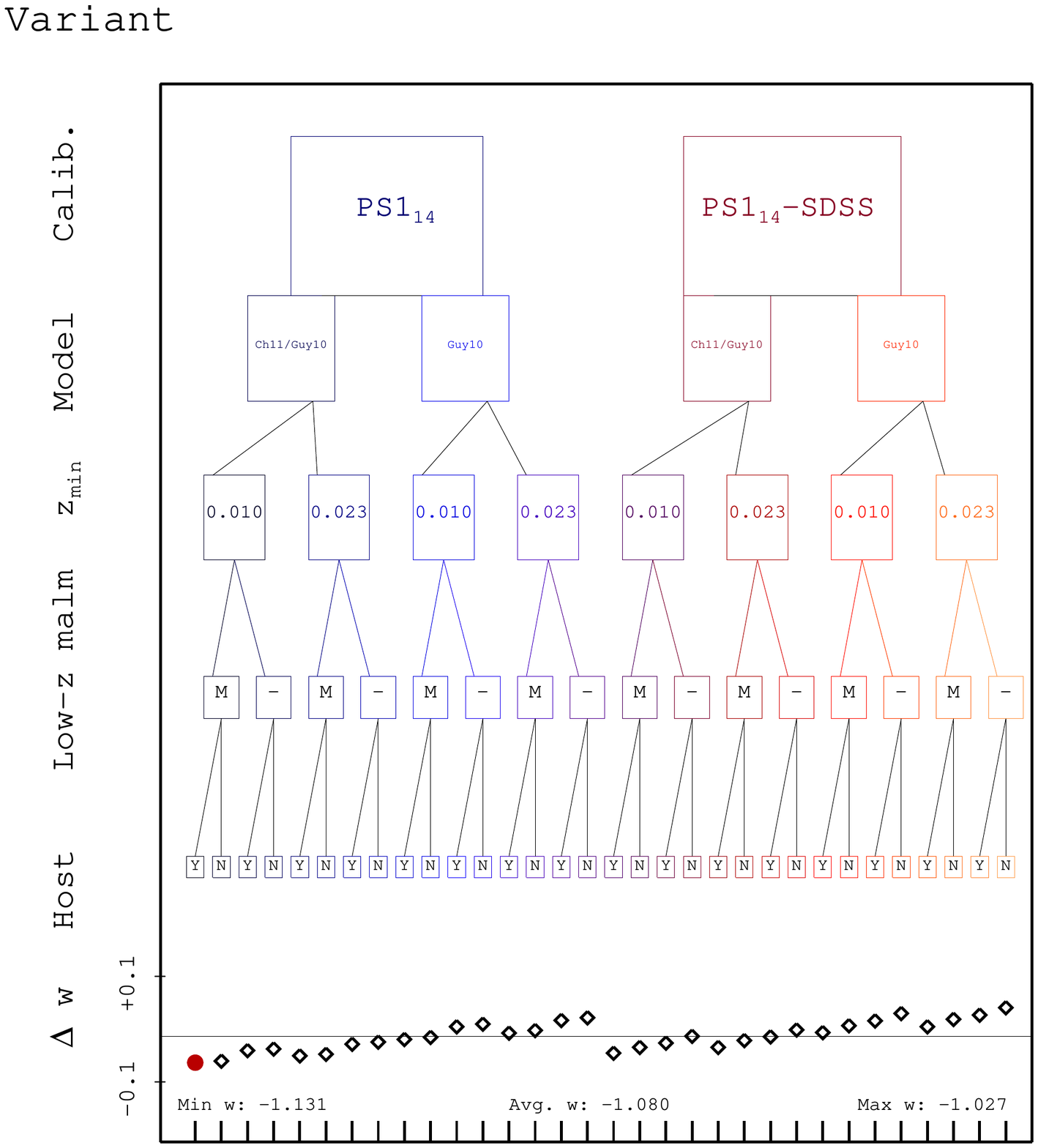}
\caption{Variants in analysis (SN+CMB+BAO+$H_0$ constraints) for the largest systematic uncertainties.  The variants are given in Table 5.  The change in $w$ is given on the bottom of the figure.  The baselines analysis is the leftmost branch.}
\label{fig:variants}
\end{figure}

\begin{figure}
\centering
\epsscale{\xxScale}  
\plotone{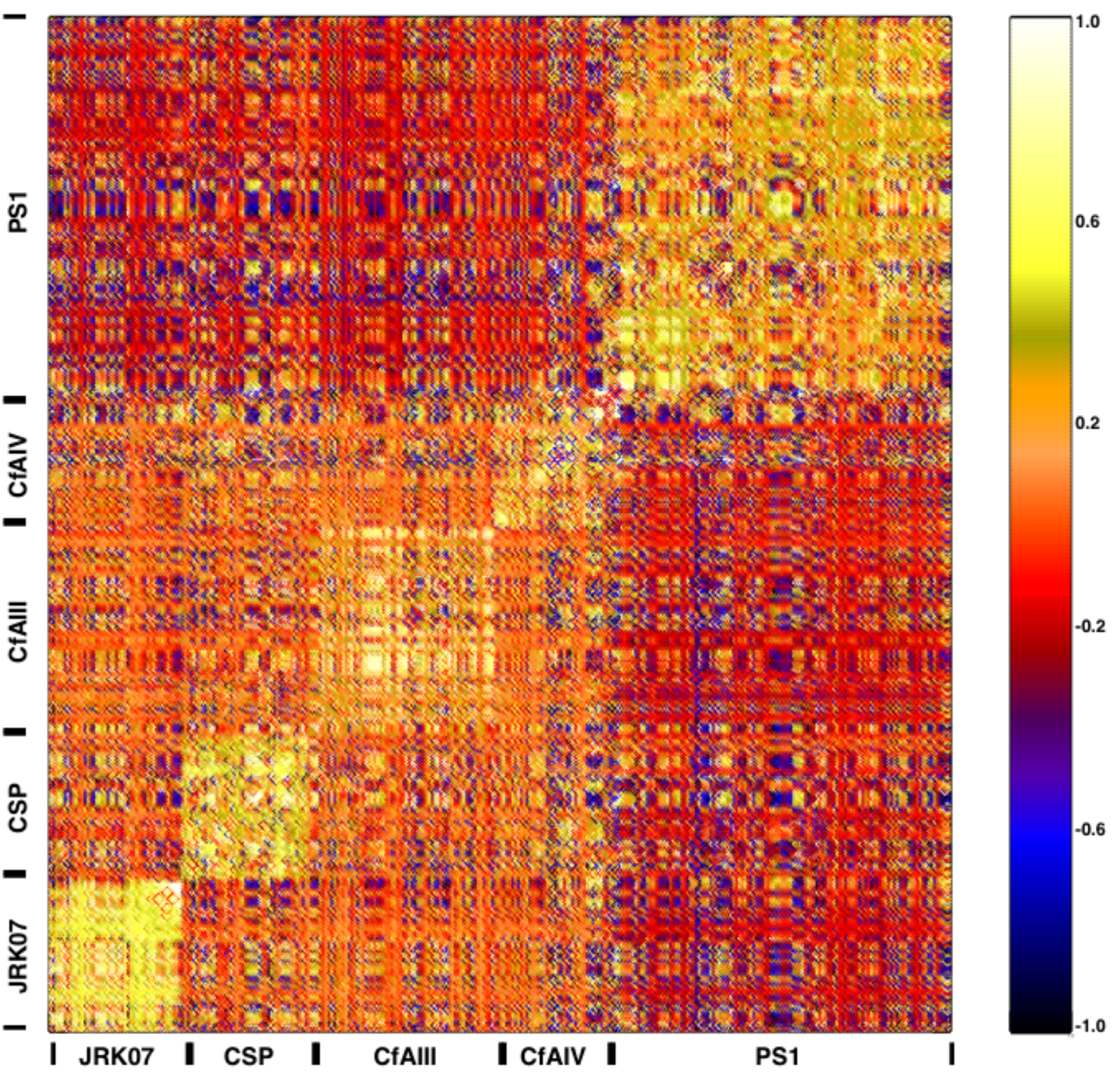}
\caption{Covariance matrix for all systematic uncertainties in the PS1+lz sample.  The matrix is sorted by sample and redshift within each sample.}
\label{fig:covar}
\end{figure}

We include results from our two methods of analyzing the dominant systematic uncertainties here.  In Table~\ref{tab:dom} we provide values for $w$, $\Omega_M$, and $\textrm{Relative Area}$ for each of the largest systematic uncertainties in the PS1+lz uncertainty budget.  The full covariance matrix for all of the systematic uncertainties in the PS1+lz sample is shown in Fig.~\ref{fig:covar} \footnote{Covariance matrix can be downloaded at http://ps1sc.org/transients/ .}.  For the covariance matrix, new values of $\alpha$ and $\beta$ are found ($\alpha=0.149,~\beta=3.15$), though they are only slightly different from those found in our statistical only case ($\alpha=0.147,~\beta=3.13$). This figure of the covariance matrix is sorted by sample and redshift to give a sense of how the uncertainties are correlated.  The difference in recovered cosmology for variants in our analysis is shown in Fig.~\ref{fig:variants}.  In this figure, all five variants discussed throughout this paper are combined and the changes in $w$ are shown.  By combined, we explicitly mean that we choose a specific series of decisions in our analysis.

From the two methods, we may understand the significance of each systematic uncertainty.  From the covariance method, we find that the flux calibration uncertainty is the largest.   It is less clear which uncertainty is the second largest as each uncertainty shifts the $w$ versus $\Omega_M$ contours and increases its area.  This problem is compounded when the constraints from SN\,Ia measurements are combined with constraints from CMB, BAO and $H_0$ measurements.  A more direct, but possibly less complete, way to understand the effects of each systematic uncertainty is from the variant method.  In Table~\ref{tab:variants}, we show the effects on $w$ when changing methods or assumptions in our analysis.  In Fig.~\ref{fig:variants}, we combine all of these variants to show the dispersion in recovered values of $w$.  These variants do not have the same magnitudes as the uncertainties stated to determine the covariance matrix.  For example, to determine the error from the calibration, we compare the PS1 and SDSS photometric systems rather than analyzing the impact of the $1\sigma$ uncertainty in each passband.  The effect of modifying the PS1 photometric system to better agree with the SDSS system is $\Delta w \approx 0.018$ (SN only).  The next largest uncertainties are due to the low-z Malmquist bias and whether we assume that intrinsic variation may be from either color-dominated or luminosity-dominated models, or only the conventional luminosity-dominated model .  

From Table 5, the total standard deviation on $w$ from these variants is $\sim0.033$ without external cosmological constraints, and $\sim0.028$ with those constraints.  This systematic uncertainty for the SN-only case is much smaller than that found from the covariance method ($\sim0.20$), as we only consider the subset of uncertainties in which we would modify the analysis.  The total systematic uncertainty after combining the external constraints is more similar, and strongly depends on how the covariance matrix shifts the recovered values of $w$ and $\Omega_M$.

\begin{deluxetable*}{l|llll}
\tablecaption{Variants in Analysis Relative to primary Method
\label{tab:variants}}
\tablehead{
\colhead{Variant} &
\colhead{$\Delta \Omega_M$} &
\colhead{$\Delta w$} &
\colhead{$\Delta \Omega_M $(+Ext.)}&
\colhead{$\Delta w$ (Ext.)} \\
}
\startdata
Nominal: & 0.0 (0.217) & 0.0 (-1.010)  & 0.0 (0.284) & 0.0 (-1.131) \\
Calib (PS1* vs PS1-SDSS): & -0.013 & 0.018  & -0.001 & -0.006 \\
Intrinsic Variation (Color/Lum.* vs. Lum.): & -0.005 & 0.044  & 0.005 & 0.027 \\
$z_{min}$ (0.01* vs. 0.023): & 0.001 & 0.013  & 0.002 & 0.011 \\
Low-z Malmquist (Yes* vs. No): & -0.007  & 0.023  & 0.001  & 0.006 \\
Host Gal. Corr. (No* vs Yes): & 0.016  & 0.003  & 0.006  & 0.026 \\

\enddata
\tablecomments{The retrieved cosmology (SN only) for variants in the analysis relative to our primary method.  The variants are: Calibration (PS1* or the SDSS modification of PS1), variation model (the average of the models with Ch11 or Guy10 scatter* or only the one with Guy10 scatter model), minimum redshift ($z=0.01$* or $z=0.023$), whether a low-z Malmquist correction is applied (Yes* or No) and whether a correction to distances based on host galaxy masses is applied (No* or Yes).  Asterisk implies primary method. Values given are e.g. $w_{+Variant}-w_{Primary}$. For the effects on $w$ from each variant shown here, all other variants follow the primary method.  Combinations of these variants are shown in Fig.~\ref{fig:variants}.}
\end{deluxetable*}

\begin{deluxetable*}{l|ccc|ccc}
\tablecaption{Overall Systematics
\label{tab:dom}}
\tablehead{
\colhead{Systematic} &
\colhead{$\Delta \Omega_M$ } &
\colhead{$\Delta w$ } &
\colhead{Rel.area}  &
\colhead{$\Delta \Omega_M$ } &
\colhead{$\Delta w$ } &
\colhead{Rel. area }  \\
~ & ~ & SN Only & ~ & ~ & SN+BAO+CMB+$H_0$ & ~ 
}
\startdata
Stat. Only & 0.000 & 0.000 & 1.000 & 0.010 & 0.000 & 1.000 \\
 & $(\Omega_M=0.223^{+0.209}_{-0.221})$ & $(w=-1.010^{+0.360}_{-0.206})$ & ~ & $(\Omega_M=0.284^{+0.010}_{-0.010})$  & ($w=-1.131^{+0.049}_{-0.049}$) & \\
\\
\hline
\\
Calibration & 0.024 & -0.093 & 1.566 & 0.006 & -0.035 & 1.309 \\
Color Model & 0.004 & -0.023 & 1.117 & 0.009 & -0.012 & 1.106 \\
Host Gal. & 0.006 & 0.000 & 1.035 & 0.011 & 0.006 & 0.977 \\
Malmquist & 0.004 & -0.011 & 1.024 & 0.010 & -0.002 & 1.015 \\
$z_{min}$ & 0.000 & -0.004 & 1.020 & 0.010 & -0.003 & 1.023 \\
MW Extinction & 0.001 & -0.007 & 1.020 & 0.010 & -0.004 & 1.019 \\
\\
\hline
\\
Syst.+Stat. & 0.032 & -0.108 & 1.697 & 0.006 & -0.035 & 1.333 \\
 & $(\Omega_M=0.256^{+0.201}_{-0.174})$ & $(w=-1.120^{+0.450}_{-0.357})$ & ~ & $(\Omega_M=0.280^{+0.013}_{-0.012})$  & ($w=-1.166^{+0.072}_{-0.069}$) & \\

\enddata
\tablecomments{The dominant systematic uncertainties in the PS1+lz sample with respect to $\Omega_m$ and $w$.  $\Omega_m$ and $w$ are given for a flat universe and $\Delta\Omega_m$ and $\Delta w$ values are the relative values such that, e.g. $w_{stat+sys}-w_{stat}$.  Individual systematic uncertainties for each of the PS1 passbands as well as the systematic uncertainties for each low-z sample.  $\textrm{RelativeArea}$ is the size of the contour that encloses 68.3\% of the probability distribution between $w$ and $\Omega_M$ compared with that of statistical only uncertainties.}
\end{deluxetable*}

We also must consider that this paper has only analyzed systematic uncertainties associated with the supernova samples, and not from the external constraints (CMB, BAO, and $H_0$ measurements).    We show in Table~\ref{tab:wmap2} how $w$ and $\Omega_M$ depend on various combinations of measurements of the different cosmological probes.  We compare the constraints on $w$ and $\Omega_M$ when we use Planck \citep{Planck13XVI} or WMAP \citep{Hinshaw+12} measurements as well as when including SNLS3 and Union2 SN measurements (\citealp{Conley_etal_2011} and \citealp{suzuki_2012_aa} respectively).  As shown in Table~\ref{tab:wmap2}, there is a difference of $\Delta w \approx -0.4$ between using Planck or WMAP data alone, and a difference of $\Delta w \approx+0.05-0.15$ when constraints from the different CMB measurements are combined with constraints from SNe measurements and/or BAO and $H_0$ measurements.  This difference would be one of the largest systematic uncertainties.

We also find that constraints on $w$ when combining PS1-lz with Planck are only $1\sigma$ from constraints when combining Union2 and Planck and $0.5\sigma$ from constraints on $w$ when combining SNLS and Planck.  While we only show combinations of up to three cosmological probes in Table~\ref{tab:wmap2}, we find that $w=-1.166^{+0.072}_{-0.069}$ when combining PS1-lz+Planck+BAO+$H_0$ and $w=-1.124^{+0.083}_{-0.065}$ when combining PS1-lz+WMAP+BAO+$H_0$.  We also note that as found by \citep{Planck13XVI}, there is $2\sigma$ tension with $w=-1$ when SNLS+Planck+BAO are combined, and $3\sigma$ tension with $w=-1$ when SNLS+Planck+$H_0$ are combined.  We therefore cannot conclude whether the tension with a flat $\Lambda$CDM model is a combination of statistical and systematic uncertainties or an implication that the model may be incorrect.

The PS1 sample analyzed in this paper is drawn from the first year and a half of the Medium Deep Survey.  At the end of the 3-year PS1 survey, the total SN\,Ia sample will be $\sim 3\times$ larger.  The increased sample size may help to improve our understanding of some of the systematic uncertainties discussed in this paper.  Relations between color, stretch, and host galaxy properties with luminosity will all likely be better understood with more SNe.  One of the main advantages to a larger sample is the ability to analyze various subsets of the data.  For example, analysis of SNe\,Ia with particularly blue colors may be especially revealing about the nature of color variation.  

To reduce the other systematic uncertainties, external data to the Medium Deep Survey will be needed.  For absolute calibration, there is still room for improvement by including more Calspec standards, particularly those that are covered by multiple surveys so that full comparisons can be done between the surveys.  This analysis will improve the accuracy in which PS1 is included into a larger sample.  Another promising development in the next year is better absolute calibration of Calspec standards (Kaiser et al. in prep).  

In this paper, we have modeled the selection effects of the PS1 survey as well as the other low-z surveys.  As more samples are combined to make one large, `Union'-like compilation, it is imperative that selection effects are publicized and standardized.  For a survey like PS1, where spectroscopic selection may have a degree of randomness, selection modeling may be inherently difficult.  We have shown here one method to analyze selection effects without an underlying knowledge of the spectroscopic follow-up program.  This method may be used for any survey, though it is possible that the method may yield biases if spectroscopic follow-up has distinct programs.

There are other uncertainties we do not include in this analysis.  Gravitational lensing may affect the apparent brightness of our supernovae, as the brightness may change via amplification or de-amplification. There is a redshift-dependent additional scatter due to this amplification of $\sigma_{\mu-\textrm{lens}}=0.055z$ \citep{2010MNRAS.405..535J}.  For the SN\,Ia included in our sample, this additional uncertainty is dominated by photometric errors and needs to be better understood.

\begin{figure*}[h!]
\centering
\epsscale{\xxScale}  
\plotone{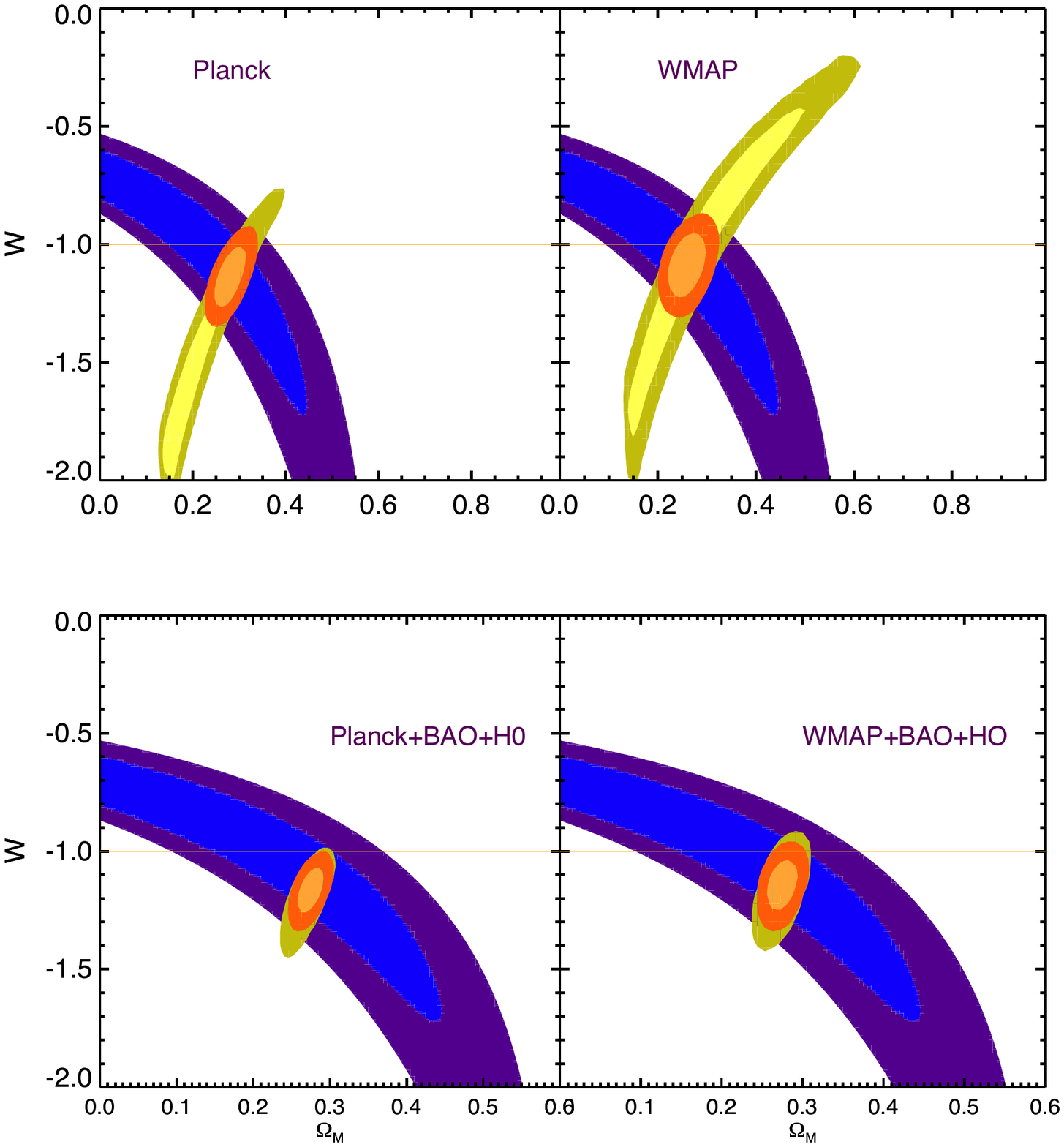}
\caption{Cosmological constraints (68\% and 95\%) from Planck and WMAP with and without constraints from SNe, BAO, and $H_0$.  Constraints from the SNe are given in Blue.  External constraints are given in Yellow and combined constraints are given in Orange.}
\label{fig:wmap}
\end{figure*}

\begin{deluxetable*}{l|c|c|c|c|c}
\tablecaption{Dependency of cosmological constraints on samples
\label{tab:wmap2}}
\tablehead{
\colhead{Measurements} &
\colhead{Planck} &
\colhead{WMAP} &
\colhead{Planck+PS1-lz} &
\colhead{WMAP+PS1-lz}   \\
~ & $\Omega_M~~~~w$  & $\Omega_M~~~~w$  & $\Omega_M~~~~w$&  $\Omega_M~~~~w$}
\startdata
- & $0.22^{0.02}_{-0.08}$ ~~~~ $-1.49^{0.25}_{-0.43}$ & $0.30^{0.07}_{-0.16}$ ~~~~ $-1.02^{0.67}_{-0.26}$ &  $ 0.28^{+0.02}_{-0.02}$  ~~~~  $-1.14^{+0.08}_{-0.08}$  &  $ 0.27^{+0.02}_{-0.02}$  ~~~  $-1.07^{+0.08}_{-0.09}$ \\
+BAO & $0.29^{0.02}_{-0.02}$ ~~~~ $-1.13^{0.14}_{-0.10}$ & $0.30^{0.02}_{-0.02}$ ~~~~ $-0.98^{0.17}_{-0.11}$ &  $ 0.29^{+0.01}_{-0.01}$  ~~~~  $-1.12^{+0.08}_{-0.07}$ &  $ 0.29^{+0.02}_{-0.02}$  ~~~  $-1.07^{+0.10}_{-0.10}$ \\
+HST & $0.26^{0.02}_{-0.02}$ ~~~~ $-1.24^{0.09}_{-0.09}$ & $0.25^{0.02}_{-0.02}$ ~~~~ $-1.14^{0.12}_{-0.10}$ &  $ 0.27^{+0.02}_{-0.02}$  ~~~~  $-1.17^{+0.06}_{-0.06}$ &  $ 0.25^{+0.01}_{-0.02}$  ~~~  $-1.10^{+0.08}_{-0.07}$ \\
+Union2 & $0.29^{0.02}_{-0.03}$ ~~~~ $-1.09^{0.09}_{-0.08}$ & $0.28^{0.02}_{-0.03}$ ~~~~ $-1.02^{0.09}_{-0.09}$ &  $ 0.29^{+0.02}_{-0.02}$  ~~~~  $-1.11^{+0.06}_{-0.06}$ &  $ 0.27^{+0.02}_{-0.02}$  ~~~  $-1.05^{+0.08}_{-0.07}$ \\
+SNLS & $0.28^{0.02}_{-0.02}$ ~~~~ $-1.13^{0.07}_{-0.07}$ & $0.26^{0.02}_{-0.02}$ ~~~~ $-1.07^{0.07}_{-0.07}$ &  $ 0.28^{+0.01}_{-0.02}$  ~~~~  $-1.13^{+0.06}_{-0.05}$  &  $ 0.26^{+0.02}_{-0.02}$  ~~~  $-1.07^{+0.06}_{-0.05}$ \\
+PS1 & $ 0.28^{+0.02}_{-0.02}$  ~~~~  $-1.14^{+0.08}_{-0.08}$  &  $ 0.27^{+0.02}_{-0.02}$  ~~~  $-1.07^{+0.08}_{-0.09}$ & - & -\\

\enddata
\tablecomments{Constraints on $\Omega_M$ and $w$ for different combinations of cosmological probes.  We assume zero curvature.  Columns 2-5 show constraints when only CMB measurements are analyzed with the measurements stated in the left-most column.  Columns 6-9 show constraints from when measurements of the CMB and PS1-lz are combined with the other probes.}
\end{deluxetable*}

The impact of the systematic uncertainties in the PS1+lz analysis on constraints of $w$ are similar to those presented by C11 and \cite{2011ApJ...737..102S} for the SNLS3+SDSS+low-z sample.  While their sample size is 472 compared to our 313, and they have many more higher-z objects from SNLS and HST, the relative area of their $\Omega_M$ versus $w$ constraints when including all of the systematic uncertainties is 1.85 compared to our 1.697.  Their systematic uncertainties are mostly dominated by calibration (relative area is 1.79 of total 1.85) while ours is 1.57 out of 1.70.  Part of the small difference in this comparison is due to our larger estimates of the uncertainties in the underlying color model.  Uncertainties due to the minimum redshift of the sample are not included in the SNLS3 final error budget, though only have a small contribution to our total uncertainty.  It is also likely that the uncertainty in our Malmquist bias correction for the PS1 sample may be larger than that for SNLS3 as our spectroscopic follow-up program was much less complete.  It is also interesting that as can be seen in Table 7,  there is a $\Delta w = -0.04$ between the PS1 and SNLS results.  From Fig. 16 we see that if we varied our baseline analysis to be more compatible with the SNLS analysis (photometric system agrees with SDSS, Guy10 model of scatter, host galaxy correction), we would shift our results by $\Delta w = +0.047$.  Therefore our results are quite similar.

Finally, we remark here that the \SNIAPSused~SN\,Ia analyzed here are a small fraction ($\sim10\%$) of all the likely SN\,Ia discovered by the Medium Deep Survey.  While larger SN\,Ia surveys are being undertaken, it is increasingly common to collect SN\,Ia light curves without a spectrum to identify the type and redshift.  Recently, \cite{campbell_13_sdss} attempted one of the first cosmological analyses of a photometrically selected SN Ia sample drawn from the full SDSS SN survey.  Although such an analysis should improve the statistical precision, the systematic uncertainties are likely to increase because of uncertainties in classification and selection biases.  Careful accounting of systematic uncertainties for an exclusively photometric sample is imperative. 

\section{Conclusions}
\label{sec:conclus}

We have presented a review of the dominant systematic uncertainties in the cosmological parameters from analysis of SN\,Ia data from the first year and a half of the PS1~Medium Deep Survey along with external low-z SN\,Ia data.  Although statistical uncertainties are still the major component of the errors in our recovered parameters, systematic uncertainties make up $>60\%$ of the relative area of the SN-only $w$ versus $\Omega_M$ constraint.   Including systematic uncertainties and assuming flatness in our analysis of only SNe measurements, we find $w=$\wSNstasys.  This result is consistent with a cosmological constant.  When including systematic uncertainties in our analysis of cosmological parameters and constraints from CMB, BAO and $H_0$ measurements, we derive~$\Omega_m=\fomCsys$~and $w=\wCsys$.  With the combined constraints, we are beginning to see either tension between the various cosmological constraints or tension with a cosmological constant.  We have shown that including systematic uncertainties in our analysis not only increases the errors on our measurements of $w$, but can shift them up to $-8.1\%$.

While the number of SN\,Ia included in this sample does not yet make the largest SN\,Ia sample, by the end of the survey it most likely will be of equal or greater size than the SDSS or SNLS samples.  The PS1 sample is already unique in the redshift range of SN\,Ia discovered ($0.03<z<0.65$).  Here, we have not only outlined how to combine the PS1 SN\,Ia sample with other samples, but also how a larger sample will help us resolve some of the issues discussed.

Our largest source of systematic uncertainty is from absolute calibration, in particular the zeropoints of the PS1 passbands.  The second largest systematic uncertainty is due to incomplete understanding of SN color.  While we correct SN distances based on the average of simulations of different models of the intrinsic scatter of SN\,Ia, choosing one model or the other changes $w$ by $\sim0.03$.

PS1 is the first of a new generation of wide-field surveys, and therefore serves as a critical test-bed for evaluating how these surveys will move SN cosmology forward.  Increasing sample sizes is no long sufficient to improve our constraints on cosmological parameters.  Detailed analysis of systematic errors, like calibration and selection efficiency, is now a critical component of any new SN cosmology survey.  Future surveys should plan their surveys in an effort to minimize these uncertainties.

\acknowledgments

\acknowledgements
{\it Facilities:}
\facility{PS1 (GPC1)},
\facility{Gemini:South (GMOS)},
\facility{Gemini:North (GMOS)},
\facility{MMT (Blue Channel spectrograph)},
\facility{MMT (Hectospec)},
\facility{Magellan:Baade (IMACS)},
\facility{Magellan:Clay (LDSS3)}
\facility{APO (DIS)}.

\acknowledgments

The Pan-STARRS1 Surveys (PS1) have been made possible through
contributions of the Institute for Astronomy, the University of
Hawaii, the Pan-STARRS Project Office, the Max-Planck Society and its
participating institutes, the Max Planck Institute for Astronomy,
Heidelberg and the Max Planck Institute for Extraterrestrial Physics,
Garching, The Johns Hopkins University, Durham University, the
University of Edinburgh, Queen's University Belfast, the
Harvard-Smithsonian Center for Astrophysics, the Las Cumbres
Observatory Global Telescope Network Incorporated, the National
Central University of Taiwan, the Space Telescope Science Institute,
the National Aeronautics and Space Administration under Grant
No. NNX08AR22G issued through the Planetary Science Division of the
NASA Science Mission Directorate, the National Science Foundation
under Grant No. AST-1238877, the University of Maryland, and Eotvos
Lorand University (ELTE).
Some observations reported here were obtained at the MMT Observatory, a joint facility of the Smithsonian Institution and the University of Arizona.
Based on observations obtained at the Gemini Observatory, which is operated by the
Association of Universities for Research in Astronomy, Inc., under a cooperative agreement
with the NSF on behalf of the Gemini partnership: the National Science Foundation
(United States), the National Research Council (Canada), CONICYT (Chile), the Australian
Research Council (Australia), Minist\'{e}rio da Ci\^{e}ncia, Tecnologia e Inova\c{c}\~{a}o
(Brazil) and Ministerio de Ciencia, Tecnolog\'{i}a e Innovaci\'{o}n Productiva (Argentina).
This paper includes data gathered with the 6.5-m Magellan Telescopes located at Las Campanas Observatory, Chile.
Based on observations obtained with the Apache Point Observatory 3.5-meter telescope, which is owned and operated by the Astrophysical Research Consortium.
Partial support for this work was provided by National Science
Foundation grant AST-1009749.
The ESSENCE/SuperMACHO data reduction pipeline {\it photpipe}
was developed with support from National Science
Foundation grant AST-0507574, and HST programs GO-10583 and GO-10903.
RPKs supernova research is supported in part by NSF Grant AST-1211196 and HST program GO-13046.
CWS and GN thank the DOE Office of Science for their support under grant ER41843.
Some of the computations in this paper were run on the Odyssey cluster
supported by the FAS Science Division Research Computing Group at
Harvard University.
This research has made use of the CfA Supernova Archive, which is
funded in part by the National Science Foundation through grant AST
0907903.
This research has made use of NASA's Astrophysics Data System.
Finally, we thank Rick Kessler for many useful discussions.  We also thank Alex Conley and Dave Jones for very helpful suggestions.

\appendix

\subsection{Photometric Calibration}

In Table~\ref{tab:calcheck2}, we present the synthetic measurements of HST standard stars.  In Table~\ref{tab:calcheck}, we present the observed magnitudes of the HST standard stars.  Differences between observed and synthetic photometry are presented in Fig.~3.  Here we give information about each HST Calspec standard, along with the files containing the HST Calspec spectra.  An important distinction is whether the spectrum is measured partly with STIS or not, and we decide not to include kf08t3 and kf01t5 because they were only observed with NICMOS.

\begin{deluxetable*}{lllllllll}
\tablecaption{Positions and Synthetic Measurements of HST Standard Stars
\label{tab:calcheck2}}
\tablehead{
\colhead{Star} &
\colhead{RA (J2000)} &
\colhead{Dec (J2000)} &
\colhead{\gps} &
\colhead{\rps} &
\colhead{\ips} &
\colhead{\zps} &
\colhead{File}  \\
~ & ~ & ~ & [Mag] & [Mag] & [Mag] & [Mag] & \\
}
\startdata
sf1615 & 244.559 & 0.00241 &  16.986 &  16.553 &  16.375 &  16.303 & sf1615\_001a\_stisnic\_003.ascii \\ 
p177d & 239.807 & 47.6116 &  13.693 &  13.307 &  13.175 &  13.147 & p177d\_stisnic\_003.ascii \\ 
wd1657 & 254.713 & 34.3151 &  16.219 &  16.685 &  17.062 &  17.355 & wd1657\_stisnic\_003.ascii \\ 
kf08t3 & 268.818 & 66.1699 &  13.641 &  13.002 &  12.733 &  12.659 & kf08t3\_nic\_001.ascii  \\ 
kf01t5 & 271.016 & 66.9286 &  14.010 &  13.259 &  12.955 &  12.865 & kf01t5\_nic\_001.ascii \\ 
kf06t2 & 269.658 & 66.7812 &  14.421 &  13.597 &  13.248 &  13.078 & kf06t2\_stisnic\_001.ascii \\ 
lds & 323.068 & 0.25408 &  14.566 &  14.798 &  15.027 &  15.231 & lds749b\_stisnic\_003.ascii \\ 
1740346 & 265.144 & 65.4542 &  12.473 &  12.521 &  12.628 &  12.726 & 1740346\_stisnic\_001.ascii \\ 
gd153 & 194.260 & 22.0310 &  13.115 &  13.582 &  13.969 &  14.255 & gd153\_stisnic\_003.ascii \\ 
c26202 & 53.1370 & -27.8633 &  16.670 &  16.359 &  16.252 &  16.232 & c26202\_stisnic\_003.ascii \\ 
gd71 & 88.1146 & 15.8879 &  12.814 &  13.260 &  13.637 &  13.925 & gd71\_stisnic\_003.ascii \\ 
wd1057 & 165.143 & 71.6342 &  14.506 &  14.822 &  15.155 &  15.632 & wd1057\_719\_stisnic\_003.ascii \\ 

\enddata
\tablecomments{ The synthetic magnitudes for the 12 calspec standard stars with adequate photometry are presented here along with the standards' locations and Calspec files used to determine synthetic values.  Most recent Calspec files are found here: http://www.stsci.edu/hst/observatory/cdbs/calspec.html.}  
\end{deluxetable*}
\begin{deluxetable*}{lllllllllllll}
\tablecaption{Standard Star Magnitudes in the PS1~System After All Corrections
\label{tab:calcheck}}
\tablehead{
\colhead{Star} &
\colhead{$m_g$} &
\colhead{$\delta m_g$} &
\colhead{$n$} &
\colhead{$m_r$} &
\colhead{$\delta m_r$} &
\colhead{$n$} &
\colhead{$m_i$} &
\colhead{$\delta m_i$} &
\colhead{$n$} &
\colhead{$m_z$} &
\colhead{$\delta m_z$} &
\colhead{$n$}  \\
~ & [Mag] & [Mag] & \# & [Mag] & [Mag] & \# & [Mag] & [Mag] & \# & [Mag] & [Mag] & \# \\
}
\startdata
sf1615 & 16.977 & 0.009 & 6 & 16.533 & 0.009 & 8 & 16.367 & 0.009 & 10 & 16.287 & 0.009 & 9 \\ 
p177d & 13.697 & 0.009 & 12 & 13.309 & 0.008 & 13 & 13.177 & 0.009 & 11 & 13.144 & 0.008 & 18 \\ 
wd1657 & 16.209 & 0.008 & 42 & 16.685 & 0.008 & 43 & 17.069 & 0.008 & 47 & 17.355 & 0.008 & 42 \\ 
kf08t3 & 13.636 & 0.011 & 2 & - & - & - & - & -& - & - & - & - \\ 
kf01t5 & 14.015 & 0.009 & 7 & 13.279 & 0.010 & 3 & - & - & - & 12.885 & 0.010 & 3 \\ 
kf06t2 & 14.416 & 0.009 & 12 & 13.591 & 0.009 & 8 & 13.251 & 0.009 & 6 & 13.086 & 0.009 & 8 \\ 
lds & 14.564 & 0.009 & 5 & 14.793 & 0.009 & 6 & 15.037 & 0.009 & 6 & - & - & - \\ 
1740346 & 12.474 & 0.009 & 6 & 12.520 & 0.009 & 7 & 12.628 & 0.009 & 6 & 12.728 & 0.009 & 8 \\ 
gd153 & - & - & - & 13.605 & 0.009 & 7 & 13.982 & 0.009 & 8 & 14.253 & 0.009 & 8 \\ 
c26202 & 16.659 & 0.008 & 304 & 16.355 & 0.008 & 304 & 16.226 & 0.008 & 303 & 16.226 & 0.008 & 312 \\ 
gd71 & 12.829 & 0.009 & 11 & 13.256 & 0.009 & 11 & 13.641 & 0.009 & 12 & 13.925 & 0.009 & 12 \\ 
wd1057 & 14.522 & 0.009 & 9 & - & - & - & - & - & - & 15.634 & 0.009 & 11 \\ 

\enddata
\tablecomments{ The 12 calspec standard stars with adequate photometry are presented here.  The uncertainties in these measurements, along with the number of observations for each standard, are shown.  The errors are given as the standard deviations of the mean.  * indicates a standard observed in \cite{Tonry12}.}
\end{deluxetable*}

\begin{figure}[h!]
\centering
\epsscale{0.5}  
\plotone{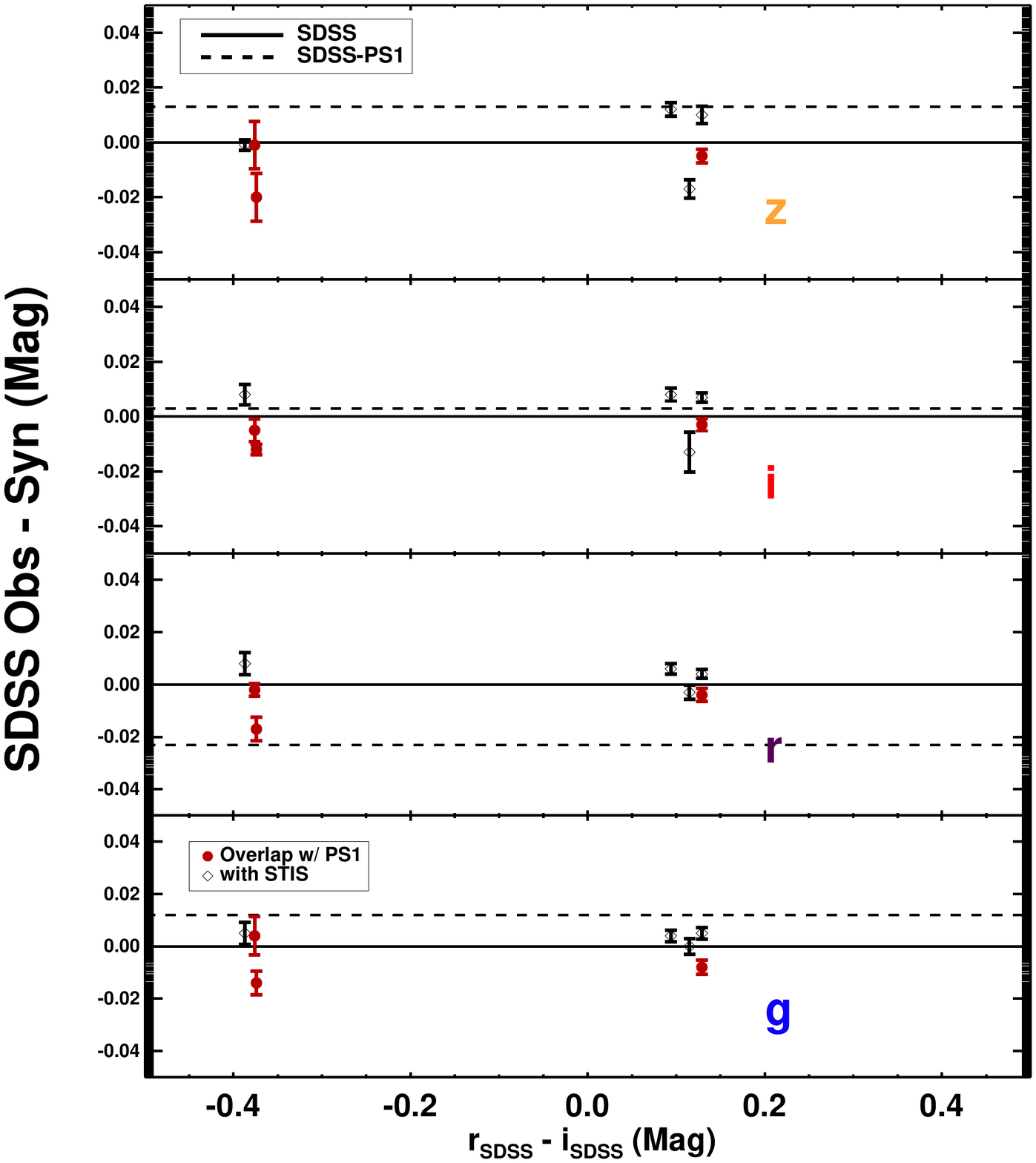}
\caption{The magnitude differences in each passband between real and synthetic SDSS observations of 7 Calspec standards. The standards observed by both PS1 and SDSS are highlighted in red.  The dashed line represents how the observed magnitudes should be shifted into agreement to better match transformations between the PS1 and SDSS calibrations.  For SDSS, the Calspec standards were measured using the SDSS PT, and have been color transformed to the SDSS 2.5m filter system.}
\label{fig:sdss_cal}
\end{figure}

To better understand the uncertainties in the PS1 implementation of the AB system, in Sec. 3 we compared photometry between SDSS and PS1.  We note that the SDSS photometric system itself may be partly the cause of discrepancies between the two systems.  In Fig.~\ref{fig:sdss_cal}, we compare observed and synthetic SDSS photometry of Calspec standards.  The photometry values are taken from \cite{Betoule2012}.
\subsection{Modeling the Selection Biases in the Low-z Sample}

\begin{figure}
\centering
\epsscale{\xScale}  
\plottwo{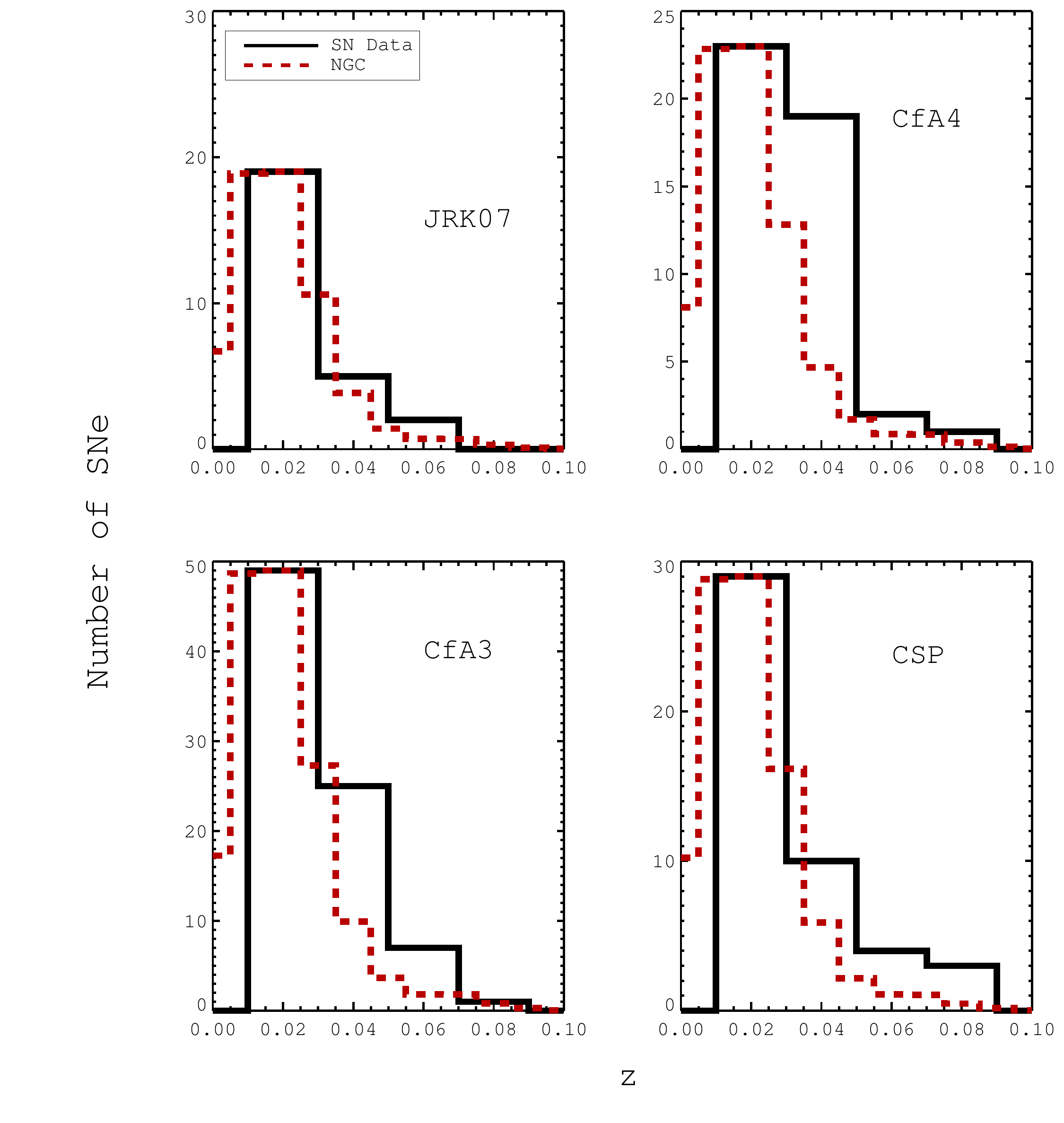}{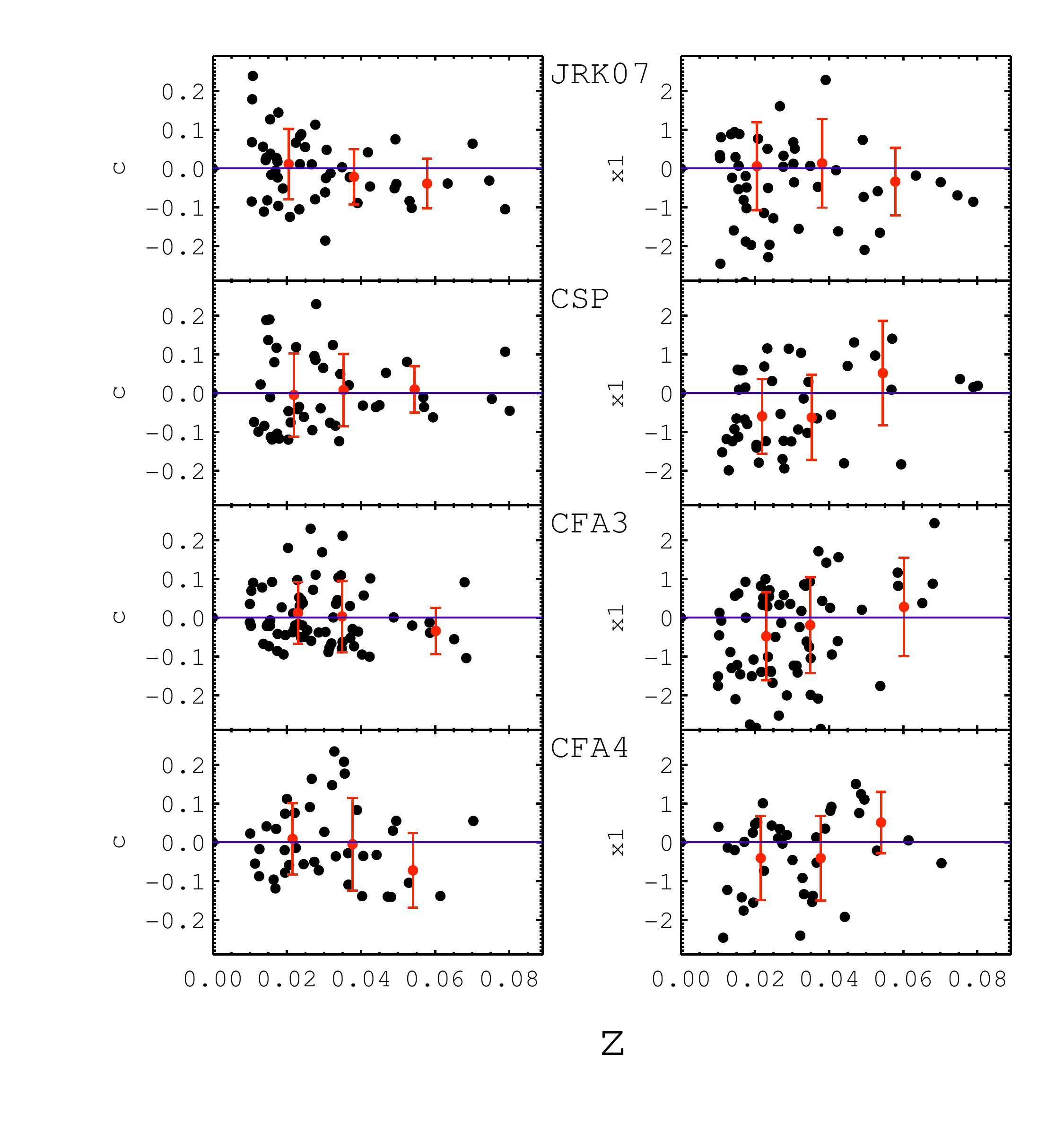}
\caption{Left: Redshift distribution of each low-z sample in comparison to the redshift distribution of the NGC catalog.  The size of the NGC distribution is scaled down to compare to the SNe sample. Right: Color and stretch versus redshift for the various low-z surveys.  A trend with redshift towards bluer colors (lower c) typically signifies a Malmquist bias.  }
\label{fig:Malmquist_low}
\end{figure}

In Fig.~\ref{fig:Malmquist_low}, we compare the redshift distribution of the low-z subsamples with the redshift distribution of the galaxies listed in the NGC.  The similarity implies that most SNe were discovered as part of galaxy-targeted surveys and that the low-z sample is not magnitude-limited.  This claim is further tested by observing the trends of color and stretch with redshift.  While a non-zero trend may be seen for CfA3 and CfA4, it is not seen for CSP or JRK07.

\end{document}